\documentclass[12pt]{article}
\usepackage{amsthm}
\usepackage{eurosym}
\usepackage{amsfonts}
\usepackage{amssymb,amsmath}
\usepackage[dvips]{color}
\usepackage{amscd}
\usepackage{tikz}
\usepackage{mathtools}
\usepackage{graphicx}
\usepackage{caption}
\usepackage{subcaption}
\usepackage[all,cmtip]{xy}

\usetikzlibrary{decorations.pathreplacing}

\usetikzlibrary{patterns}

\def\and{\mathrm{and}}

\newtheorem{lemm}{Lemma}
\newtheorem{prop}{Proposition}

\newtheorem{thm}{Theorem}

\newtheorem{remark}{Remark}
\newcommand{\be}{\begin{equation}}
\newcommand{\ee}{\end{equation}}
\newcommand{\bea}{\begin{eqnarray}}
\newcommand{\eea}{\end{eqnarray}}
\newcommand{\beas}{\begin{eqnarray*}}
\newcommand{\eeas}{\end{eqnarray*}}
\newcommand{\ba}{\begin{array}}
\newcommand{\ea}{\end{array}}

\newcommand{\nbox}{{\,\lower0.9pt\vbox{\hrule \hbox{\vrule height 0.2 cm \hskip 0.19 cm \vrule height 0.2 cm}\hrule}\,}}

\def\href#1#2{#2}
\begin{document}

\begin{titlepage}
\hfill
\vbox{
    \halign{#\hfil         \cr
           } 
      }  

\hbox to \hsize{{}\hss \vtop{ \hbox{}

}}

%

\vspace*{20mm}

\begin{center}

{\large \textbf{Calabi-Yau generalized complete intersections and aspects of\\ \vspace{3mm} cohomology of sheaves}}

{\normalsize \vspace{10mm} }

{\normalsize {Qiuye Jia${}^{1,2}$, Hai Lin${}^{1}$}  }

{\normalsize \vspace{10mm} }

{\small \emph{${}^1$\textit{Yau Mathematical Sciences Center, Tsinghua University,
Beijing 100084, P. R. China
}} }

{\normalsize \vspace{0.2cm} }

{\small \emph{$^2$\textit{Department of Mathematical Sciences, Tsinghua University,
Beijing 100084, P. R. China
\\
}} }

{\normalsize \vspace{0.4cm} }

\end{center}

\begin{abstract}

We consider generalized complete intersection manifolds in the product space of projective spaces, and work out useful aspects pertaining to the cohomology of sheaves over them. First, we present and prove a vanishing theorem on the cohomology groups of sheaves for subvarieties of the ambient product space of projective spaces. We then prove an equivalence between configuration matrices of complete intersection Calabi-Yau manifolds. We also present a formula of the genus of curves in generalized complete intersection manifolds. Some of these curves arise as the fixed point locus of certain symmetry group action on the generalized complete intersection Calabi-Yau manifolds. We also make a blowing-up along the curves, by which one can generate new Calabi-Yau manifolds. Moreover, an approach on spectral sequences is used to compute Hodge numbers of generalized complete intersection Calabi-Yau manifolds and the genus of curves therein.

\end{abstract}

\end{titlepage}

\vskip 1cm

\section{Introduction}

\label{sec_Introduction}\vspace{1pt}\renewcommand{\theequation}{1.%
\arabic{equation}} \setcounter{equation}{0}

\vspace{1pt}

The complete intersection Calabi-Yau manifolds \cite%
{Candelas:1987kf,Green:1986ck,Yau1985} might be a competitive candidate for
spacetime model in string theory. In this method of constructing a
Calabi-Yau manifold, the complicated Calabi-Yau geometry is embedded into a
relatively simple ambient space, such as a product of projective spaces.
Many geometric quantities of the Calabi-Yau manifolds can be deduced by
their relations to those of the ambient space. In this paper, we explore a
generalization of complete intersection Calabi-Yau manifolds \cite%
{Anderson:2015iia}. These complete intersection Calabi-Yau (CICY) manifolds,
including three-folds \cite{Candelas:1987kf,Green:1986ck,Yau1985} and
four-folds \cite%
{Brunner:1996bu,Gray:2013mja,Gray:2014kda,Gray:2014fla,Anderson:2016cdu},
have been constructed and investigated extensively.

\vspace{1pt}

Such intersections can be described in the language of line bundles. The
complete intersection Calabi-Yau manifolds are defined by the common zero
locus of polynomials, which are global sections of line bundles with
non-negative degrees. This can be generalized to the construction of new
Calabi-Yau manifolds, through bundles with no global sections on the ambient
space \cite{Anderson:2015iia}. In this new construction, negative degrees
are allowed, that is, one can replace polynomials of non-negative degrees by
rational functions. These manifolds are solutions to systems of algebraic
equations in a product of projective spaces where the functions in the
defining equations may have negative degrees. This is to allow the existence
of poles. The Laurent polynomials in these equations have poles, but these
poles avoid the common intersection locus, and this largely expands the
construction of complete intersection Calabi-Yau varieties \cite%
{Anderson:2015iia,Berglund:2016yqo,Berglund:2016nvh}.

\vspace{1pt}

However, we do not require all the line bundles to have global sections at
the first place, and we need to take intersections step by step \cite%
{Anderson:2015iia}, to construct these generalized complete intersection
Calabi-Yau (gCICY) varieties. The idea is to take a submanifold $M$ in the
ambient manifold $A$ and to consider submanifolds $X$ in $M$. These
submanifolds $X$ need not be complete intersections in the ambient manifold $%
A$. To be more specific, the domain that we require these sections to be
regular is decreasing everytime when we take intersections. Hence, although
the line bundle does not have a global section on the entire product of
projective spaces, it has regular section when restricted to appropriate
subvarieties. One then constructs the generalized complete intersection
Calabi-Yau in these subvarieties.

\vspace{1pt}

In this paper we work on aspects of cohomology of sheaves over generalized
complete intersection Calabi-Yau manifolds. We will develop some tools and
approaches, in order to understand Calabi-Yau manifolds better. First, we
introduce and prove a generalized version of vanishing theorem on the
cohomology groups of sheaves that will be used in our computation of Hodge
numbers of the gCICYs. This vanishing theorem is a generalization of the
original theorem \cite{Deligne1,Katz Sarnak} from the case of a single
projective space to the case of a product of several projective spaces. In
the process of proving this generalized vanishing theorem, we used Poincare
residue exact sequences and the method of induction. This cohomology group
in the vanishing theorem will appear in the double complexes and the long
exact sequences of the cohomology groups that we are interested in computing.

\vspace{1pt}

When we consider certain symmetry group actions on gCICYs, some curves would
be the fixed point locus. We make a blow up of the generalized complete
intersection Calabi-Yau manifolds along these curves which we identified as
the fixed point locus of some involutions. We compute the genus of curves
which themselves can be viewed as generalized complete intersection
manifolds in a product of projective spaces, and work out a general formula
of their genus. This genus formula is needed for the Hodge numbers of the
blow ups.

\vspace{1pt}

We devote a spectral sequence approach to the computation of Hodge numbers
of gCICYs. These Hodge numbers not only encode topological information, but
also give the dimensions of the moduli spaces. We show that an approach on
spectral sequences can be used to efficiently compute the topological data
of the generalized complete intersection manifolds, including the one
dimensional case of curves. We build spectral sequences of double complexes
of cohomology groups of sheaves and compute their dimensions. Moreover, one
of the methods that we obtain the data of an appropriately twisted sheaf is
to tensor known exact sequences by locally free sheaves. We also used this
method when we prove the generalized vanishing theorem.

\vspace{1pt}

We also prove a birational equivalence between configuration matrices of
complete intersection Calabi-Yau manifolds. This equivalence relation is
slightly different from the one propsed in \cite{Candelas:1987kf}, which is
instead a diffeomorphic equivalence. This equivalence relation would be
useful in the classification of generalized complete intersection Calabi-Yau
manifolds \cite{Anderson:2015iia}.

\vspace{1pt}

The gCICYs would give us manifolds that didn't arise in the exhaustive list
of CICYs. In addition, an important phenomenon in the deformation of CICYs
is that in general, the moduli space of CICYs has dimension less than the
moduli space of total deformation class. This is easy to understand: the
manifolds that can be defined by polynomials are only a part among all those
CY manifolds in this deformation class. It is conceivable that, after
generalizing to gCICYs, we would increase the dimension of the moduli
spaces. This fact coincides with the result that some of gCICYs are not
previously constructed in the list of CICYs. In understanding their
properties, we can make use of modern tools in sheaf theory and cohomology
theory. Moreover, gCICYs is a promising candidate of important physics model
\cite{Anderson:2015iia,Berglund:2016yqo,Berglund:2016nvh}. In conjunction
with the characterization of moduli spaces, we are able to understand CY
manifolds better.

\vspace{1pt}

The organization of this paper is as follows. In Section \ref{sec_intro_},
we briefly review the idea behind the generalization of CICY, and introduce
the necessary background. In Section \ref{sec_generalized_vanishing_theorem}%
, we propose a vanishing theorem on the cohomology groups of sheaves for
subvarieties of the ambient product space of projective spaces, with our
identified condition. And then in Section \ref%
{sec_identity_configuration_matrix}, we prove an equivalence between
configuration matrices of complete intersection Calabi-Yau manifolds.
Afterwards in Section \ref{sec_genus_formula}, we consider involution of
gCICY and the fixed point locus which are curves. We present a general
formula of the genus of the curves in gCICY. We also make a blow up along
the curves. In Section \ref{sec_spectral_sequence_approach}, we present a
spectral sequence approach of the cohomology groups of sheaves of gCICY,
which are used to compute Hodge numbers and genus of curves, among other
things. Finally, we discuss our results and draw some conclusions in Section %
\ref{sec_discussion}. In Appendix \ref{appendix_special_case}, we include
some details pertaining to a special case of the genus formula. In Appendix %
\ref{appendix_topological_data}, additional computational details of Hodge
numbers and other topological data are included. In this paper, a variety is
assumed to be a scheme of finite type over a field $k$, where $k$ is the
complex number field in all the sections but it can be any algebraically
closed field in Section \ref{sec_generalized_vanishing_theorem}.

\vspace{1pt}


\section{Generalized Complete Intersection Manifolds and \newline
Calabi-Yau Manifolds}

\label{sec_intro_}

\renewcommand{\theequation}{2.\arabic{equation}} \setcounter{equation}{0} %
\renewcommand{\thethm}{2.\arabic{thm}} \setcounter{thm}{0} %
\renewcommand{\theprop}{2.\arabic{prop}} \setcounter{prop}{0}

\vspace{1pt}

We devote this section to emphasize the idea behind the generalization of
CICY \cite{Anderson:2015iia}. We will explore the construction of
generalized complete intersection Calabi-Yau manifolds given in \cite%
{Anderson:2015iia}. These manifolds are constructed by line bundles on an
ambient space. Here the ambient space $A~$is a product of projective spaces,
\begin{equation}
A=\mathbb{P}^{n_{1}}\times \mathbb{P}^{n_{2}}\times \cdots \times \mathbb{P}%
^{n_{t}}.
\end{equation}

Let us consider a line bundle $\mathcal{L}_{a}$ on $A$, with multi-degree ($%
q_{a}^{1},\cdots ,q_{a}^{t}$), that is%
\begin{equation}
\mathcal{L}_{a}=O_{A}(q_{a}^{1},\cdots ,q_{a}^{t})=\otimes _{i}\pi
_{i}^{\ast }O_{\mathbb{P}^{n_{i}}}(q_{a}^{i}).  \label{line_bundle_01}
\end{equation}%
Here $\pi _{i}$ is the projection of $A$ onto each projective space factor, $%
O_{\mathbb{P}^{n_{i}}}(q_{a}^{i})$ is a line bundle on each projective
space, and the $a$ labels each line bundle.

Let $M$ be a submanifold of $A$ that is defined by the common zero locus of
several polynomial sections of line bundles on $A$. The degrees of these
line bundles can be described by a matrix, which can be called the
configuration matrix. Now we can add more columns in the configuration
matrix. $X$ is the submanifold of $M$ that is defined by those sections of
bundles on $M$ that correspond to those additional columns of the
configuration matrix which contain negative entries. In other words, we can
write them using configuration matrices as follows:
\begin{equation}
M=%
\begin{matrix}
\mathbb{P}^{n_{1}} \\
\mathbb{P}^{n_{2}} \\
\vdots \\
\mathbb{P}^{n_{t}}%
\end{matrix}%
\begin{bmatrix}
q_{1}^{1} & \cdots & q_{P}^{1} \\
q_{1}^{2} & \cdots & q_{P}^{2} \\
\vdots &  & \vdots \\
q_{1}^{t} & \cdots & q_{P}^{t}%
\end{bmatrix}%
\subset A.  \label{M}
\end{equation}%
\begin{equation}
X=%
\begin{matrix}
\mathbb{P}^{n_{1}} \\
\mathbb{P}^{n_{2}} \\
\vdots \\
\mathbb{P}^{n_{t}}%
\end{matrix}%
\begin{bmatrix}
q_{1}^{1} & \cdots & q_{P}^{1} & \cdots & q_{K}^{1} \\
q_{1}^{2} & \cdots & q_{P}^{2} & \cdots & q_{K}^{2} \\
\vdots &  &  &  & \vdots \\
q_{1}^{t} & \cdots & q_{P}^{t} & \cdots & q_{K}^{t}%
\end{bmatrix}%
\subset M.  \label{X}
\end{equation}

In the above, each column $(q_{a}^{1},\cdots ,q_{a}^{t})^{T}$ of a
configuration matrix corresponds to a line bundle $\mathcal{L}_{a}$ with
multi-degree ($q_{a}^{1},\cdots ,q_{a}^{t}$), as defined in Eq. (\ref%
{line_bundle_01}). The $a$ labels each column, and equivalently, each line
bundle. The integers themselves denote the degrees of the defining
polynomials in the homogeneous coordinates of the projective space factors.

The matrix elements of the configuration matrix (\ref{M}) of the ordinary
complete intersection $M$ are non-negative integers. That is, $q_{a}^{i}\geq
0$, for $a=1,\cdots ,P$. Each column of the configuration matrix of $M$,
defines a codimension-one hypersurface in the ambient space $A$. If there
are $P$ columns, then $M$ is the complete intersection of $P$ hypersurfaces
in $A$. Each constraint is defined by a polynomial equation, in which the
polynomial is a section $s_{a}\in H^{0}(A,\mathcal{L}_{a})\neq 0,\forall
a=1,\cdots ,P$. Each polynomial is a homogenous polynomial with multi-degree
($q_{a}^{1},\cdots ,q_{a}^{t}$) in the homogenous coordinates of the $t$
projective space factors. $M$ is a subvariety in $A$ with codimension $P$,
because all the degrees of the line bundles, defining $M$, as a complete
intersection, are non-negative.

The original construction of complete intersection Calabi-Yau manifolds did
not allow negative integers in the configuration matrix, or in order words,
negative degrees for the line bundles \cite%
{Candelas:1987kf,Green:1986ck,Yau1985}. By taking into account
non-polynomial deformations in the moduli space of Calabi-Yau varieties,
construction of generalized complete intersection Calabi-Yau have been
proposed \cite{Anderson:2015iia}. The generalization is that one allows
negative degrees for the line bundles, or in other words, negative integers
in the configuration matrix. The additional columns, which are the last $K$$%
- $$P$ columns in the configuration matrix (\ref{X}) of $X$, each contain
negative integers. We have that for $\forall a>P,$ $\exists $ $q_{a}^{i}<0$
in the line bundles defined in Eq. (\ref{line_bundle_01}), in which case
these line bundles do not have global sections on $A$. These additional
columns with negative integers correspond to the algebraic equations in
products of projective spaces involving rational functions that have
negative degrees. The pole locus of these rational functions avoid the
common intersection locus $M$. Denoting the pole locus of these rational
functions on $A$ to be $\Delta \subset A$, we require that $M\cap \Delta
=\varnothing $.

Below, we illustrate the reason to consider these line bundles with no
global sections on the ambient space and sketch the process to take
intersections step by step \cite{Anderson:2015iia}. Because there exists
certain negative degree in the line bundle $\mathcal{L}_{a}$, for $a>P$,
associated to the additional column in the configuration matrix of $X$, we
have that $H^{0}(A,\mathcal{L}_{a})=0$ and there is no global section of
this line bundle over $A$. As a consequence, $X$ is not a submanifold in $A$
defined by the global sections of line bundles on $A$. Let's consider a set
of subvarieties $M_{a},~a=P+1,\cdots ,K,~$inside $M$, and $M_{a}\subset
M_{a-1}$. And let us denote $M_{P}=M$. The $M_{a}$ is a subvariety in $%
M_{a-1}$ defined by the section $s_{a}\in H^{0}(M_{a-1},\mathcal{L}%
_{a}|_{M_{a-1}})\neq 0$ for $a=P+1,\cdots ,K$. We take intersections step by
step. The domain that we require the section to be regular is decreasing
everytime we take intersection. Finally, $X=M_{K}$, which is a submanifold
in $M_{K-1}$. So, although the line bundle does not have a global section on
the entire product of projective spaces, it has sections that is regular on
the subvarieties described above. If we consider those restrictive
conditions one by one, say, from the left to the right in the configuration
matrix (\ref{X}), rather than simultaneously, we would obtain global
sections. Hence $X$ has codimension $K$$-$$P$ in $M$. In sum, we have $X%
\overset{i_{X}}{\rightarrow }M\overset{i_{M}}{\rightarrow }A$, where $i_{X}$
is the inclusion of $X$ in $M$, and $i_{M}$ is the inclusion of $M$ in $A$,
and we denote $i=i_{M}\circ i_{X}$.

We now turn to the condition for the above configuration matrix to represent
a Calabi-Yau manifold. Since $X$ is a submanifold of $M$, and $M$ is in turn
a submanifold of $A$, we use the adjunction formulas iteratively, see the
versions in Appendix B. We hence use the adjunction formulas of the tangent
sheaves in the sequence form iteratively to obtain the Calabi-Yau condition
for $X$. Denote the dual of the ideal sheaf of $X$ to be $\mathcal{E}$. We
use the fact that $\mathcal{E}=\oplus _{a}\mathcal{E}_{a},\mathcal{E}%
_{a}=i^{\ast }O_{A}(q_{a}^{1},\cdots ,q_{a}^{t})$, where $q_{a}^{i}$ are the
matrix elements in the configuration matrix, $a=1,2,\cdots ,K$. In our
process of computing the Chern classes, we use the fact that since $\mathcal{%
E}$ is line bundle, its Chern class $c_{k}$ will vanish as long as $k\geq 2$%
, hence we have%
\begin{equation}
c(\mathcal{E})=\prod_{a}c(\mathcal{E}_{a})=\prod_{a}(1+%
\sum_{i}q_{a}^{i}J_{i}),
\end{equation}%
where $J_{i}=c_{1}(i^{\ast }\pi _{i}^{\ast }O_{\mathbb{P}^{n_{i}}}(1))$. The
dimension of the $i$-th projective space factor is $n_{i}$. Hence using the
adjunction formulas of the tangent sheaves iteratively, we have $%
c_{1}(X)=\sum_{i}(n_{i}+1-\sum_{a}q_{a}^{i})J_{i}$.~Hence the condition on
the matrix elements of the configuration matrix for a Calabi-Yau manifold is
\begin{equation}
\sum_{a}q_{a}^{i}=n_{i}+1,  \label{a1}
\end{equation}%
where $q_{a}^{i}$ here can be either non-negative or negative. The above
Calabi-Yau condition is that the sum of the matrix elements of each row
equals the dimension of the projective space of the corresponding row plus
1. In the case of ordinary CICY, $q_{a}^{i}$ are non-negative only.

We can determine the Hodge numbers of the manifolds and then use them as
topological data to classify those manifolds. Among other things, the Hodge
numbers will give us the complex structure deformation space of the family
of manifolds. To be more precise, configuration matrices describe a family
of manifolds and a manifold is only determined after choosing its complex
structure within that family. We are particularly interested in three-folds
and four-folds for the manifolds $X$ due to their usefulness in
compactification of string theory.

\vspace{1pt}

\section{Generalization of a Vanishing Theorem}

\label{sec_generalized_vanishing_theorem}

\renewcommand{\theequation}{3.\arabic{equation}} \setcounter{equation}{0} %
\renewcommand{\thethm}{3.\arabic{thm}} \setcounter{thm}{0} %
\renewcommand{\theprop}{3.\arabic{prop}} \setcounter{prop}{0}

\vspace{1pt}

The original vanishing theorem on the vanishing of the cohomology group $%
H^{0}(Y,T_{Y})$ for a hypersurface $Y$ in a single projective space, was
based on the works of Bott, Deligne, Kodaira-Spencer, Matsumura-Monsky and
overviewed in \cite{Katz Sarnak}. We are going to generalize it to the case
of the product of projective spaces. When we consider complete intersection
or generalized complete intersection manifolds, we want to consider the
product of projective spaces to be the ambient space. The vanishing theorem
that we generalized is useful for computing the Hodge numbers of the
generalized complete intersection manifolds. This cohomology group would
appear in the double complexes and the long exact sequences of the
cohomology groups that we are interested in computing. We will denote the
product by
\begin{equation}
A=\prod_{i=1}^{t}\mathbb{P}^{n_{i}+1}.
\end{equation}%
Although the results we have in other sections in this paper will only be
the analytic case, which specializes the field to be the complex number
field, our result in this section also holds for schemes over other
algebraically closed fields.

\subsection{Preparation Using Sheaves}

Before presenting the generalized theorem and diving into the details, we
develop some general results on divisors and sheaves. The ambient space $%
A=\prod_{i=1}^{t}\mathbb{P}^{n_{i}+1}$, as the product space of projective
spaces, have multi-dimension $n+1=(n_{1}+1,n_{2}+1,\cdots ,n_{t}+1)$. Note
that here $n$ is a vector $(n_{1},n_{2},\cdots ,n_{t})~$with $t$ components.
We also consider the ambient space $A$ to be a smooth scheme over an
algebraically closed field $k$ and denote it as $A/k$. The specific case we
are going to treat is a subvariety $Y~$of multi-degree $d=(d_{1},d_{2},%
\cdots ,d_{t})$, which is defined by the zero of a homogenous polynomial of
degrees $d_{1},d_{2},\cdots ,d_{t}$ respectively in the homogenous
coordinates of the $t$ projective space factors of $A$.

For brevity, for $a=(a_{1},a_{2},\cdots ,a_{t})$, we denote the line bundle
as follows, $O(a)=\otimes _{i}\pi _{i}^{\ast }O_{\mathbb{P}%
^{n_{i}+1}}(a_{i}) $.$~$We define the sheaves%
\begin{equation}
E(a)=E(a_{1},\cdots ,a_{t})=E\otimes O(a)=E\otimes \pi _{1}^{\ast }O_{%
\mathbb{P}^{n_{1}+1}}(a_{1})\otimes \cdots \otimes \pi _{t}^{\ast }O_{%
\mathbb{P}^{n_{t}+1}}(a_{t}),
\end{equation}%
when tensoring a given sheaf $E$ with $O(a)~$on $A$, where $\pi _{i}$
denotes the projection onto the $i$-th factor of $A$. The similar notation
will also be used when we restrict the domains to be subschemes of $A$. For $%
m=(m_{1},m_{2},\cdots ,m_{t})$, we denote the sheaves%
\begin{equation}
\Omega ^{m}(a)=\pi _{1}^{\ast }\Omega _{\mathbb{P}^{n_{1}+1}}^{m_{1}}\otimes
\cdots \otimes \pi _{t}^{\ast }\Omega _{\mathbb{P}^{n_{t}+1}}^{m_{t}}\otimes
O(a),
\end{equation}%
and in the case when all the $m_{i}$ are the same, we simply denote the
vector $m$ as $m=(m,m,\cdots ,m)$.

The canonical bundle of $A$ is $K_{A}=O_{A}(-(n+2))$. This is a short-hand
notation for $O_{A}(-(n_{1}+2),-(n_{2}+2),\cdots ,-(n_{t}+2))$. The
canonical bundle of $Y$ is%
\begin{equation}
K_{Y}=(K_{A}\otimes I_Y ^{\ast })|_{Y},
\end{equation}%
where $I_Y $ is the ideal sheaf of $Y$. We have that $I_Y =O_{A}(-d)$, hence
the canonical bundle of $Y$ is $K_{Y}=O_{Y}(d-n-2)$. The tangent bundle is $%
~T_{Y/k}=\Omega _{Y/k}^{n-1}\otimes K_{Y/k}^{-1}$, and hence
\begin{equation}
T_{Y/k}=\Omega _{Y/k}^{n-1}(n+2-d).
\end{equation}%
We are interested in computing $H^{p}(Y,T_{Y/k})$, which is useful for and
facilitates the determination of the Hodge numbers of the generalized
complete intersection manifolds.

Let's discuss some general results on divisors. Consider the ambient space $%
A/k$ as a smooth scheme, and $Y$ is a divisor which is smooth as a scheme
over $k~$\cite{Katz Sarnak}. We denote $Der_{Y}(A/k)$ to be a subsheaf of $%
T_{A/k}~$that is characterized by the property that it sends the ideal sheaf
$I_{Y}$ to itself. Now $Der_{Y}(A/k)$ is a locally free $O_{A}$-module. We
denote the twisted sheaf of forms, defined as the dual of $Der_{Y}(A/k)$, by
$\Omega _{A/k}^{1}(\log Y)$, which is also locally free. There is a perfect
pairing induced by the contraction of forms along vector fields
\begin{equation}
\Omega _{A/k}^{1}(\log Y)\times Der_{Y}(A/k)\rightarrow O_{A},
\end{equation}%
hence we have that%
\begin{equation}
\Omega _{A/k}^{1}(\log Y)={\mathrm{Hom}}_{O_{A}}(Der_{Y}(A/k),O_{A}).
\end{equation}%
Since we have $Der_{Y}(A/k)\subset T_{A/k}$, there is a reversed inclusion
of their duals $\Omega _{A/k}^{1}\subset \Omega _{A/k}^{1}(\log Y)$. For $%
q\geq 0$, we define $\Omega _{A/k}^{q}(\log Y)=$ $\wedge ^{q}(\Omega
_{A/k}^{1}(\log Y))$.

We have the Poincare residue map, for $q\in \mathbb{Z}_{>0}$,%
\begin{equation}
\varphi :\Omega _{A/k}^{q}(\log Y)\longrightarrow \Omega _{Y/k}^{q-1},
\end{equation}%
which is a surjective map and $\ker \varphi ~$is $\Omega _{A/k}^{q}$, and
for brevity in the above notation, we have used the brief notation $\Omega
_{Y/k}^{q-1}$ to mean $i_{\ast }(\Omega _{Y/k}^{q-1})$, where $%
i:Y\rightarrow A~$is the inclusion. We will use this brief notation
throughout the paper. We hence have the following two short exact sequences.
The Poincar\'{e} residue exact sequence is
\begin{equation}
0\longrightarrow \Omega _{A/k}^{q}\longrightarrow \Omega _{A/k}^{q}(\log
Y)\longrightarrow \Omega _{Y/k}^{q-1}\longrightarrow 0.  \label{21}
\end{equation}%
The restriction exact sequence is
\begin{equation}
0\longrightarrow I_Y \otimes \Omega _{A/k}^{q}(\log Y)\longrightarrow \Omega
_{A/k}^{q}\longrightarrow \Omega _{Y/k}^{q}\longrightarrow 0.  \label{22}
\end{equation}%
The ideal sheaf is $I_Y =O_{A}(-d)$. We stick to the convention that $\Omega
^{q}=0$ whenever $q<0$.

\subsection{Statement of the Theorem and Proof by Induction}

Now we present and prove a vanishing theorem on the cohomology groups of
sheaves for subvarieties of the ambient product space of projective spaces.

\begin{thm}
\label{generalized_vanishing_theorem} Suppose $n_{i}\geq 0$, $%
\sum_{i=1}^{t}d_{i}\geq 2t+1+s,d_{i}\geq 1$, where $s$ is the number of $%
n_{i}$ which are 1. The ambient space is the product space of projective
spaces $A=\prod_{i=1}^{t}\mathbb{P}_{k}^{n_{i}+1}$. Then for an
algebraically closed field $k$, and a smooth hypersurface $Y/k$ of
multi-degree $d=(d_{1},d_{2},\cdots ,d_{t})$ in $A$, we have $%
H^{0}(Y,T_{Y/k})=0$.
\end{thm}

\textit{Proof.} The idea to prove the above main result is to consider an
even more general form of this equation and prove it by descending
induction. $A$ is over an algebraically closed field $k$. Since $%
T_{Y/k}=\Omega _{Y/k}^{n-1}(n+2-d)$, it is the case $p=0$ of the following
proposition $C(p)$:
\begin{equation}
C(p):\quad H^{p}(Y,\Omega _{Y}^{n-1-p}(n+2-(p+1)d))=0.
\end{equation}%
We aim to prove this equality. When we add $p~$to, or subtract $p$ from a
vector, we mean $p$ is a vector $(p,p,\cdots ,p)$, and when we do it with a
number, we view $p$ as a number.

The technique we use is to twist the short exact sequences. Next we consider
the long exact sequences of the cohomology groups of sheaves associated with
them and prove the vanishing theorems on those related sheaves altogether.
The ideal sheaf $I_{Y}$ of $Y$ in $A$ is $O_{A}(-d)$. We will repeatedly use
the following two exact sequences. Tensoring the restriction exact sequences
(\ref{22}) with $O_{A}(n+2-(p+1)d)$, we have:
\begin{equation}
\begin{array}{lr}
0\longrightarrow \Omega _{A}^{n-1-p}(\log Y)(n+2-(p+2)d))\longrightarrow
\Omega _{A}^{n-1-p}(n+2-(p+1)d)) &  \\
\longrightarrow \Omega _{Y}^{n-1-p}(n+2-(p+1)d))\longrightarrow 0. &
\end{array}
\label{short_sequence_01}
\end{equation}%
Tensoring the Poincare residue exact sequences (\ref{21}) with $%
O_{A}(n+2-(p+1)d)$, we have:
\begin{equation}
\begin{array}{lr}
0\longrightarrow \Omega _{A}^{n-p}(n+2-(p+1)d))\longrightarrow \Omega
_{A}^{n-p}(\log Y)(n+2-(p+1)d)) &  \\
\longrightarrow \Omega _{Y}^{n-1-p}(n+2-(p+1)d))\longrightarrow 0. &
\end{array}
\label{short_sequence_02}
\end{equation}

We will prove another three equalities at the same time. We list them below:
\begin{equation}
A(p):\quad H^{p}(A,\Omega _{A}^{n-p}(\log Y)(n+2-(p+1)d))=0.
\end{equation}%
And two equalities on the ambient space:%
\begin{equation}
B(p):\quad H^{p}(A,\Omega _{A}^{n-1-p}(n+2-(p+1)d))=0.  \label{26}
\end{equation}%
\begin{equation}
D(p):\quad H^{p}(A,\Omega _{A}^{n-p}(n+2-(p+1)d))=0.
\end{equation}

After taking the long exact sequences of the cohomology groups of sheaves in
the above short exact sequences, the groups in these four equalities above
are related. After taking the long exact sequence of the cohomology groups
associated to the short exact sequence (\ref{short_sequence_01}), this will
let us know: If $B(p)$ and $A(p+1)$ both hold, $C(p)$ shall hold. The long
exact sequence of the cohomology groups associated to the short exact
sequence (\ref{short_sequence_02}) will let us know another implication: If $%
C(p)$ and $D(p)$ both hold, $A(p)$ shall hold.

Now let us explain the strategy to prove these four equalities. The $A(p)$
will hold for big $p$ for $p>n_{i}$, by definition. If $B(p)$ and $D(p)$
hold for all $p$, then by the above derivation, we have that, \emph{once }$%
A(p+1)$\emph{\ holds then this implies that }$C(p)$ \emph{and then }$A(p)$%
\emph{\ would hold as well.} We are then on the stage of descending
induction from $p+1$ to $p$, which will prove that $A(p)$ and $C(p)$ hold
for all $p$.

We only need to prove $B(p)$ and $D(p)$ now. We prove $D(p)$ then: By K\"{u}%
nneth formula, we know
\begin{equation}
H^{p}(A,\Omega _{A}^{n-p}(n+2-(p+1)d))=\oplus
_{\sum_{i=1}^{t}q_{i}=p}\otimes _{i}H^{q_{i}}(\mathbb{P}^{n_{i}+1},\Omega
^{n_{i}-p}(n_{i}+2-(p+1)d_{i})).
\end{equation}

We reformulate the Bott formula as follows. $H^{\alpha }(\mathbb{P}%
^{n+1},\Omega _{\mathbb{P}^{n+1}}^{\beta }(\gamma ))$ will be nonzero only
when:\newline
(1) $\alpha =0$, and either $\gamma >\beta \geq 0$ or $\gamma =\beta =0;$%
\newline
(2) $1\leq \alpha \leq n,\alpha =\beta ,\gamma =0;$\newline
(3) $\alpha =n+1,\beta \geq 0$ and either $\beta -\gamma >n+1$ or $\beta
=n+1,\gamma =0$.

We are going to show that every tensor product that arises in the sum will
vanish. Otherwise, suppose there is a nonzero $\otimes _{i}H^{q_{i}}(\mathbb{%
P}^{n_{i}+1},\Omega ^{n_{i}-p}(n_{i}+2-(p+1)d_{i}))$. Consider the index $%
\beta $ of $\Omega ^{\beta }(\gamma )$, it needs to be non-negative, hence $%
p\leq n_{i},\forall i$. Notice that $\sum_{i}q_{i}=p$, we know that $%
q_{i}\leq n_{i}$, hence the last case (3) of non-vanishing will never happen.

If the first case (1) happens, we have $q_{i}=0$. If the second condition is
$\gamma >\beta \geq 0$, we know that $n_{i}-(p+1)d_{i}>n_{i}-p$, or $%
p>(p+1)d_{i}$, which is impossible. If the second condition reads $\gamma
=\beta =0$, we know that $n_{i}=p,n_{i}+2=(p+1)d_{i}$, enforcing $%
p+2=(p+1)d_{i}$. This can happen only when $p=0,d_{i}=2$, which makes $n_{i}$
to be 0, leading to contradiction.

Hence every factor in the tensor product has to be in the second case (2),
giving us:%
\begin{eqnarray}
&&1\leq q_{i}\leq n_{i}. \\
&&n_{i}-p=q_{i}. \\
&&n_{i}+2=(p+1)d_{i}.
\end{eqnarray}%
Summing these equations over $i$, we get $\sum_{i}[(p+1)d_{i}-p-2]=p$.~So
with the condition
\begin{equation}
\sum_{i}d_{i}\geq 2t+1+s,  \label{condition}
\end{equation}%
we have that%
\begin{equation}
p\geq (1+t+s)p+1+s.
\end{equation}%
This leads to contradiction. Hence we have proved $D(p)$.

Then we prove $B(p)$: By K\"{u}nneth formula, we know
\begin{equation}
H^{p}(A,\Omega _{A}^{n-p-1}(n+2-(p+1)d))=\oplus
_{\sum_{i=1}^{t}q_{i}=p}\otimes _{i}H^{q_{i}}(\mathbb{P}^{n_{i}+1},\Omega
^{n_{i}-p-1}(n_{i}+2-(p+1)).
\end{equation}%
For the same reason as the $D(p)$ case, the third non-vanishing case will
never happen. And we have:
\begin{equation}
p+1\leq n_{i}~~~\mathrm{and}~~~q_{i}\leq p.
\end{equation}

If the first case arises, we will have $q_{i}=0$. Moreover, once the second
condition specializes to be $\gamma >\beta \geq 0$, we have $%
n_{i}+2-(p+1)d_{i}>n_{i}-p-1$, or $(p+1)(d_{i}-1)<2$, so this enforces $%
d_{i}=1,\forall i$. This then implies that $t=\sum_{i}d_{i}\geq 2t+1+s$,
leading to contradiction. Hence the second condition must be $\gamma =\beta
=0$. This means $n_{i}=p+1,n_{i}+2=(p+1)d_{i}$. Hence $p+3=(p+1)d_{i}$. All $%
d_{i}$ will be the same. Since $\sum_{i}d_{i}>2t$, we know $d_{i}\geq 3$.
This once again enforces $p=0,d_{i}=3,n_{i}=1$. This shows $s=t$, hence $%
3t=\sum_{i}d_{i}\geq 2t+1+s$, which is contradictory again.

Now we are left with the second case once again. The equations are
\begin{equation}
n_{i}-p-1=q_{i},~~~~~n_{i}+2=(p+1)d_{i}.
\end{equation}%
Summing them over $i$ and comparing these two equations, we have
\begin{equation}
\sum_{i}n_{i}=pt+p+t=(p+1)(\sum_{i}d_{i})-2t.  \label{relation_07}
\end{equation}%
With the condition (\ref{condition}), we have that
\begin{equation}
0\geq (p-1)t+1+s(p+1).  \label{relation_08}
\end{equation}%
From this we know $p=0$, since there would be contradiction if $p\geq 1$.
Thus by the first of Eq. (\ref{relation_07}) we know that all $n_{i}$ have
to be 1 and $s=t$. Then Eq. (\ref{relation_08}) leads to contradiction. With
condition (\ref{condition}), we have showed that the non-vanishing is not
possible. Hence we have proved $B(p)$. Therefore by the descending induction
above, the theorem is completely proved. {\hfill $\square $}\newline


In the generalized vanishing theorem above, there is an interesting
condition $\sum_{i}d_{i}\geq 2t+1+s$. We will make comments on this in the
next subsection. Our proof above is for general $n_{i}~\geq 0$, with the
condition stated in the theorem.

This result will be useful in our computation of Hodge numbers of the
generalized CICYs, as in our approach in Sections \ref{sec_genus_formula}
and \ref{sec_spectral_sequence_approach}, as well as in Appendix \ref%
{appendix_topological_data}. Further, it is related to the properness of
morphisms of schemes.

Our result is also useful for many other models. The theorem we have proved
is valid not only for the examples we scrutinize in Sections \ref%
{sec_genus_formula} and \ref{sec_spectral_sequence_approach}, but also for
many other models, such as the family $X_{m}$ constructed in \cite%
{Anderson:2015iia,Berglund:2016yqo,Berglund:2016nvh}, which we recall the
definition here:%
\begin{equation}
X_{m}=%
\begin{matrix}
\mathbb{P}^{4} \\
\mathbb{P}^{1}%
\end{matrix}%
\begin{bmatrix}
1 &  & 4 \\
m &  & 2-m%
\end{bmatrix}%
,\quad F_{m}=%
\begin{matrix}
\mathbb{P}^{4} \\
\mathbb{P}^{1}%
\end{matrix}%
\begin{bmatrix}
1 \\
m%
\end{bmatrix}%
.  \label{manifolds_iii}
\end{equation}

\subsection{Version for Single Projective Space and Relation to Proper
Morphism and Haar Measure}

Our theorem \ref{generalized_vanishing_theorem} above is a generalization of
the theorem of \cite{Deligne1,Katz Sarnak}. The original version for a
single projective space is as follows.

\begin{thm}
\label{single_projective_space} Suppose $n\geq 0$, $d\geq 3$, and that $%
(n,d)\neq (1,3)$. The ambient space is a single projective space $\mathbb{P}%
_{k}^{n+1}$.$~$Then for any field k, and a smooth hypersurface $M/k$ of
degree $d~$in $\mathbb{P}_{k}^{n+1}$, we have $H^{0}(M,T_{M/k})=0$.
\end{thm}

This original version was proved in \cite{Deligne1,Katz Sarnak} for any
field $k$, in scheme-theoretic formulations. See also \cite{Deligne2} for
related discussion.

The theorem \ref{single_projective_space} for an algebraically closed field $%
k$, is implied by our theorem \ref{generalized_vanishing_theorem}, as our $%
t=1$ case. In theorem \ref{single_projective_space}, when the dimension is $%
n=1$, the lower bound of the degree $d$ have to be strengthened to be $4$.
In our theorem \ref{generalized_vanishing_theorem}, for $t=1$, when $n=1$
which means $s=1$, our condition $\sum_{i}d_{i}\geq 2t+1+s$ reduces to the
condition $d\geq 2+1+1=4$, which is in agreement with the condition in
theorem \ref{single_projective_space}. So our generalized theorem
characterizes the lower bound of the degree in a more uniform way. The
original excluded case $(n,d)\neq (1,3)$ would now be described in a more
uniform and natural formulation.

The vanishing of $H^{0}(M,T_{M})$ in theorem \ref{single_projective_space},
is related to the projective automorphism of $M$. The projective
automorphism of a hypersurface $M$ in $\mathbb{P}^{n+1}~$is the automorphism
that is induced by the automorphism of the ambient space $\mathbb{P}^{n+1}$.$%
~M$ is a scheme over a field $k$, and we denote it by $M/k$. We have its
projective automorphism%
\begin{equation}
\mathrm{ProjAut}(M/k)\subset PGL_{n+2}(k).
\end{equation}%
Suppose $H_{n,d}$ is a degree $d$ hypersurface in $\mathbb{P}^{n+1}$, which
plays the role of aforementioned $M$, see \cite{Matsumura Monsky}. The
following proposition by \cite{Mumford Fogarty} is on the properness of the
action of the group scheme $PGL_{n+2}$ on $M$.

\begin{prop}
Suppose $n\geq 0$, $d\geq 3$. The~action of the group scheme $PGL_{n+2}$ on
degree $d$ hypersurface $H_{n,d}~$in $\mathbb{P}^{n+1}$is proper and finite,
i.e., the morphism%
\begin{eqnarray}
PGL_{n+2}\times _{\mathbb{Z}}H_{n,d} &\rightarrow &H_{n,d}\times _{\mathbb{Z}%
}H_{n,d}  \notag \\
(g,h) &\mapsto &(h,g(h))
\end{eqnarray}%
is a proper and finite morphism.
\end{prop}

The properness of this action of the group scheme is in turn useful for the
Haar measure of the matrix integrals for $PGL_{n+2}(k)$ for a field $k$ \cite%
{Katz Sarnak}. Since $PGL_{n+2}$ is from the action on a single projective
space, it would be interesting to see whether there is a connection between
the generalized vanishing theorem for a product of projective spaces and
properties of the action of the product $\prod_{i}PGL_{n_{i}+2}$.

\section{Equivalence between Configuration Matrices and its Proof}

\label{sec_identity_configuration_matrix}

\renewcommand{\theequation}{4.\arabic{equation}} \setcounter{equation}{0} %
\renewcommand{\thethm}{4.\arabic{thm}} \setcounter{thm}{0} %
\renewcommand{\theremark}{4.\arabic{remark}} \setcounter{remark}{0}

In this section, we prove an equivalence between configuration matrices for
complete intersection Calabi-Yau manifolds. This equivalence means that the
manifolds defined by the configuration matrices in the two sides are
birationally equivalent. For the convenience of discussion below, we
introduce the notation $a=(a_{2},\cdots ,a_{k}),b=(b_{2},\cdots ,b_{k})$.
The indices start with 2 because of the columns that they start from. P.
Candelas et. al. proposed the following identification of CY manifolds
without proof \cite{Candelas:1987kf}, and we will give the proof of this
equivalence.

\begin{thm}
Let there be two smooth complete intersection Calabi-Yau manifolds, defined
by the following two configuration matrices respectively, then they are
birationally equivalent and have the same Hodge numbers, i.e.,%
\begin{equation}
\begin{matrix}
\mathbb{P}^{1} \\
\mathbb{P}^{n} \\
X%
\end{matrix}%
\begin{bmatrix}
1 &  & a \\
1 &  & nb \\
0 &  & {\tilde{M}}%
\end{bmatrix}%
\overset{Birational}{=}%
\begin{matrix}
\mathbb{P}^{1} \\
\mathbb{P}^{n-1} \\
X%
\end{matrix}%
\begin{bmatrix}
a+b \\
nb \\
{\tilde{M}}%
\end{bmatrix}%
.  \label{identity}
\end{equation}%
Here, $a$ and $b$ are multi-columns, ${\tilde{M}}$ is a block matrix, and $0$
is a column of zeros.
\end{thm}

\textit{Proof.} We first show how to use Calabi-Yau condition to reduce the
number of possible cases to a reasonable extent. Recall that the Calabi-Yau
condition is: the sum of a row corresponds to $\mathbb{P}^{n_{i}}$ is $%
n_{i}+1$. We have that $a=(a_{2},\cdots ,a_{k}),b=(b_{2},\cdots ,b_{k})$.$~$%
Set $F:=\sum_{i=2}^{k}a_{i},G:=\sum_{i=2}^{k}b_{i}$. The Calabi-Yau
condition is
\begin{equation}
F+1=2,\quad nG+1=n+1.
\end{equation}%
So we have $F=1,G=1$.~Since all $a_{i},b_{i}$ are non-negative integers,
this means that there is one and only one non-zero term in the $a$-row and $%
b $-row respectively. So, the part of the matrix that involves $\mathbb{P}%
^{1}\times \mathbb{P}^{n}$ non-trivially would only have at most 3 columns,
depending on whether the non-zero terms are at the same column or not.

Hence we are left with two general cases:%
\begin{equation}
\begin{matrix}
\mathbb{P}^{1} \\
\mathbb{P}^{n} \\
X%
\end{matrix}%
\begin{bmatrix}
1 &  & 1 \\
1 &  & n \\
0 &  & {\tilde{M}}%
\end{bmatrix}%
\overset{Birational}{=}%
\begin{matrix}
\mathbb{P}^{1} \\
\mathbb{P}^{n-1} \\
X%
\end{matrix}%
\begin{bmatrix}
2 \\
n \\
{\tilde{M}}%
\end{bmatrix}%
.  \label{identity_ca_04}
\end{equation}
\begin{equation}
\begin{matrix}
\mathbb{P}^{1} \\
\mathbb{P}^{n} \\
X%
\end{matrix}%
\begin{bmatrix}
1 &  & 0 &  & 1 \\
1 &  & n &  & 0 \\
0 &  & {\tilde{M}}_{1} &  & {\tilde{M}}_{2}%
\end{bmatrix}%
\overset{Birational}{=}%
\begin{matrix}
\mathbb{P}^{1} \\
\mathbb{P}^{n-1} \\
X%
\end{matrix}%
\begin{bmatrix}
1 &  & 1 \\
n &  & 0 \\
{\tilde{M}}_{1} &  & {\tilde{M}}_{2}%
\end{bmatrix}%
.  \label{identity_ca_03}
\end{equation}%
Here, ${\tilde{M}}_{i}$ and ${\tilde{M}}$ are column vectors.

We can introduce the notion of `Nef Complete Intersection'. This stands for
those complete intersections whose canonical bundle is a nef line bundle.
This is a generalization of Calabi-Yau condition. This is equivalent to the
condition $d_{i}\geq n_{i}+1$, where $d_{i}$ is the sum of the $i$-th row in
the matrix.

Now we show the birational equivalence. From now on, we use $%
([x_{0},x_{1}],[y_{0},...,y_{n}],z)=(x,y,z)$ as coordinates of $\mathbb{P}%
^{1}\times \mathbb{P}^{n}\times X$. The first case is:%
\begin{equation}
\begin{matrix}
\mathbb{P}^{1} \\
\mathbb{P}^{n} \\
X%
\end{matrix}%
\begin{bmatrix}
1 & 1 \\
1 & n \\
0 & {\tilde{M}}%
\end{bmatrix}%
.
\end{equation}%
The manifold $M_{1}$, corresponding to the first column, could in general be
defined by $x_{0}L_{0}(y)+x_{1}L_{1}(y)=0$, where $L_{i}$ has degree 1 in $y$%
. Then after a change of variable, it could be defined by $%
x_{0}y_{0}+x_{1}y_{1}=0$. Then the manifold $M_{2}$, corresponding to the
second column, could be defined by $x_{0}P_{0}(y,z)+x_{1}P_{1}(y,z)=0$,
where $P_{i}$ has degree $n$ in $y$.

Then $M_{1}$ is birationally equivalent to $\mathbb{P}^{1}\times \mathbb{P}%
^{n-1}\times X$ by the map:
\begin{equation}
E_{1}:([x_{0},x_{1}],[y_{0},...,y_{n}],z)\rightarrow
([x_{0},x_{1}],[y_{1},...,y_{n}],z),  \label{E_1}
\end{equation}%
between the region $(A_{0}\times \mathbb{P}^{n}\times X)\cap M_{1}$ and $%
A_{0}\times \mathbb{P}^{n-1}\times X$, where $A_{0}:=\{[x_{0},x_{1}]\in
\mathbb{P}^{1}:x_{0}\neq 0\}$. This is because on this region we have the
inverse of $E_{1}$:
\begin{equation}
E_{2}:([x_{0},x_{1}],[y_{1},...,y_{n}],z)\rightarrow ([x_{0},x_{1}],[-\frac{%
x_{1}}{x_{0}}y_{1},y_{1}...,y_{n}],z).
\end{equation}%
Now consider $M=M_{1}\cap M_{2}$, which is defined by
\begin{equation}
x_{0}y_{0}+x_{1}y_{1}=0,\ \ \ \ \ x_{0}P_{0}(y,z)+x_{1}P_{1}(y,z)=0.
\end{equation}

Then we deform $P_{i}$ to let all of their terms' degree in the first
variable $y_{0}$ to be at most 1, and throughout this deformation, we
require the coefficient of terms having zero-th order in $y_{0}$ to be non-zero and keep
generic with respect to the remaining $y_{i}$, so that the
polynomial can not be factorized, and hence there will not be normal
crossing singularities caused by degeneration of $P_{i}$. The family that we
obtained by changing the coefficients of the polynomials is a smooth family
of projective manifolds, in particular they form a smooth family of K\"{a}%
hler manifolds. By Proposition 9.20 in \cite{Voisin}, the Hodge numbers will
stay as constants when the change of coefficients of the polynomials are
small. Then we can cover the path, from the initial polynomial to the
polynomial having degree at most 1 in the first variable $y_{0}$, by such
neighborhoods and then pick a finite subcover using compactness of this
path. Consequently, the Hodge numbers will stay as constants throughout this
deformation.

The part where $E_{1}$ is not defined should satisfy $x_{0}=0$, $x_{1}\neq 0$%
. This is equivalent to:
\begin{equation}
x_{0}=0,\ \ \ y_{1}=0,\ \ \ P_{1}(y,z)=0.
\end{equation}%
We denote this manifold by $X_{1}$, which is $%
\begin{matrix}
\mathbb{P}^{n-1} \\
X%
\end{matrix}%
\begin{bmatrix}
n \\
{\tilde{M}}%
\end{bmatrix}%
$, because the $\mathbb{P}^{1}$ factor is now a single point and $y_{1}=0$
gives a $\mathbb{P}^{n-1}\subset \mathbb{P}^{n}$. And we set $%
Z_{1}=M\backslash X_{1}$, which is open in $M$.

We consider $Z_{2}=E_{1}(Z_{1})$ and its closure. On the region $x_{0}\neq 0$%
, we know that $E_{1}(Z_{1})$ is defined by:
\begin{equation}
x_{0}P_{0}(-\frac{x_{1}}{x_{0}}y_{1},y_{1},...,y_{n},z)+x_{1}P_{1}(-\frac{%
x_{1}}{x_{0}}y_{1},y_{1},...,y_{n},z)=0.
\end{equation}%
On the region $x_{0}\neq 0$, multiplying $x_{0}$ would result in an
equivalent condition, which is:
\begin{equation}
G(x_{0},x_{1};y_{1},...,y_{n},z):=x_{0}P_{0}(-x_{1}y_{1},y_{1},...,y_{n},z)+x_{1}P_{1}(-x_{1}y_{1},y_{1},...,y_{n},z)=0.
\end{equation}

The $G$ has bidegree $(2,n)$ in $x$ and $y$. Let $\tilde{Z}_{2}$ be the
manifold defined by $G=0$. It is the union of $Z_{2}$ and $\tilde{Z}_{2}\cap
\{x_{0}=0\}$. The intersection $\tilde{Z}_{2}\cap \{x_{0}=0\}\ $is
equivalently defined by $x_{0}=0,P_{1}(-x_{1}y_{1},y_{1},y_{2},...,y_{n})=0$%
, which is a codimension-1 submanifold of $\tilde{Z}_{2}$. And $E_{2}$ is
defined outside the codimension-1 region $\tilde{Z}_{2}\cap \{x_{0}=0\}$.
Hence $E_{1},E_{2}$ give a birational equivalence. Collecting above results
together, we have shown:
\begin{eqnarray}
&&%
\begin{matrix}
\mathbb{P}^{1} \\
\mathbb{P}^{n} \\
X%
\end{matrix}%
\begin{bmatrix}
1 &  & 1 \\
1 &  & n \\
0 &  & {\tilde{M}}%
\end{bmatrix}%
\xrightarrow{Deform}%
\begin{matrix}
\mathbb{P}^{1} \\
\mathbb{P}^{n} \\
X%
\end{matrix}%
\begin{bmatrix}
1 &  & 1 \\
1 &  & n \\
0 &  & {\tilde{M}}%
\end{bmatrix}%
\overset{Birational}{=}%
\begin{matrix}
\mathbb{P}^{1} \\
\mathbb{P}^{n-1} \\
X%
\end{matrix}%
\begin{bmatrix}
2 \\
n \\
{\tilde{M}}%
\end{bmatrix}%
.  \notag \\
&&
\end{eqnarray}

Now we consider the second case. The three columns correspond to:
\begin{align}
& x_{0}y_{0}+x_{1}y_{1}=0,  \notag \\
& P(y,z)=0,  \notag \\
& x_{0}Q_{0}(z)+x_{1}Q_{1}(z)=0.
\end{align}%
We still deform $P$ to let it has degree 1 in the first variable $y_{0}$,
which would not change the Hodge numbers, for the same reason as the first
case. Here $P$ has degree $n$ in $y$. Denote this manifold by $M$. And still
set $Z_{1}=M\cap (A_{0}\times \mathbb{P}^{n}\times X)$. Then $E_{1}(Z_{1})$
is characterized by
\begin{equation}
x_{0}\neq 0,\ \ \ P(-\frac{x_{1}}{x_{0}}y_{1},y_{1},...,y_{n})=0,\ \ \
x_{0}Q_{0}(z)+x_{1}Q_{1}(z)=0.
\end{equation}

Using $E_{1}$ in (\ref{E_1}), and also multiplying the second equation by $%
x_{0}$, we obtain a manifold $\tilde{Z}_{2}$ defined by
\begin{equation}
P(-x_{1}y_{1},y_{1},...,y_{n})=0,\ \ x_{0}Q_{0}(z)+x_{1}Q_{1}(z)=0.
\end{equation}%
And we know $\tilde{Z}_{2}\backslash (E_{1}(Z_{1}))=\tilde{Z}_{2}\cap
\{x_{0}=0\}$. The later intersection is defined by
\begin{equation}
P(-x_{1}y_{1},y_{1},...,y_{n})=0,\ \ \ Q_{1}(z)=0,\ \ \ x_{0}=0.
\end{equation}%
This is a codimension 1 submanifold of $\tilde{Z}_{2}$, outside which $E_{2}$
is defined. Hence $E_{1},E_{2}$ give a birational equivalence. That is:%
\begin{eqnarray}
&&%
\begin{matrix}
\mathbb{P}^{1} \\
\mathbb{P}^{n} \\
X%
\end{matrix}%
\begin{bmatrix}
1 & 0 & 1 \\
1 & n & 0 \\
0 & {\tilde{M}}_{1} & {\tilde{M}}_{2}%
\end{bmatrix}%
\xrightarrow{Deform}%
\begin{matrix}
\mathbb{P}^{1} \\
\mathbb{P}^{n} \\
X%
\end{matrix}%
\begin{bmatrix}
1 & 0 & 1 \\
1 & n & 0 \\
0 & {\tilde{M}}_{1} & {\tilde{M}}_{2}%
\end{bmatrix}%
\overset{Birational}{=}%
\begin{matrix}
\mathbb{P}^{1} \\
\mathbb{P}^{n-1} \\
X%
\end{matrix}%
\begin{bmatrix}
1 & 1 \\
n & 0 \\
{\tilde{M}}_{1} & {\tilde{M}}_{2}%
\end{bmatrix}%
.  \notag \\
&&
\end{eqnarray}

The two sides are birationally equivalent to each other. Moreover, they are
both Calabi-Yau, which means that both sides have trivial canonical bundles.
Hence, we can use Corollary 2.9 of \cite{Ito}, which shows that two smooth
projective birationally equivalent varieties having nef canonical bundles
have the same Hodge numbers, to conclude that both sides have the same Hodge
numbers, completing our proof. {\hfill $\square $}\newline

\begin{remark}
The method except the first step of using Calabi-Yau condition actually hold
in a more general context. The notion of `nef' is not needed in this
context. However this is a good generalization of Calabi-Yau condition,
which is much more restrictive. And by the result of Kontsevich \cite%
{Kontsevich} and Ito \cite{Ito}, Hodge numbers are birational invariants
between nef varieties. Moreover, the nef condition in the complete
intersection case can be easily characterized by replacing the equality in
the Calabi-Yau condition by an inequality $d_{i}\geq n_{i}+1$, with $d_{i}$
the sum of the $i$-$th$ row: using the adjunction formula, the canonical
bundle of a hypersurface defined by a degree $d=(d_{1},...,d_{N})$
polynomial is $\oplus _{i}O(d_{i}-n_{i}-1)$. When $d_{i}=n_{i}+1$, the
corresponding line bundle is trivial; When $d_{i}>n_{i}+1$, the bundle in
that factor would be ample, which implies that it is nef. Hence `nef' should
be a useful concept to consider when we investigate the Hodge numbers of
complete intersections.
\end{remark}

\section{Genus Formula of Curves in Generalized Complete Intersections and
Blow Up}

\label{sec_genus_formula} \renewcommand{\theequation}{5.\arabic{equation}} %
\setcounter{equation}{0} \renewcommand{\thethm}{5.\arabic{thm}} %
\setcounter{thm}{0} \renewcommand{\theprop}{5.\arabic{prop}} %
\setcounter{prop}{0}


\subsection{The Configuration Matrix and Fixed Point Locus}

\label{sec_fixed_point_locus_subsection}

We are interested in the involutions of the generalized complete
intersection Calabi-Yau manifolds $X$, whose fixed point locus are curves.
For convenience, we denote the generalized complete intersection manifold by
$X$ and the fixed point locus by $Z$. The configuration matrix of $X$ is:
\begin{equation}
X=%
\begin{matrix}
\mathbb{P}^{m} \\
\mathbb{P}^{n}%
\end{matrix}%
\begin{bmatrix}
a_{1} &  & a_{2} &  & \cdots &  & a_{p} \\
b_{1} &  & b_{2} &  & \cdots &  & b_{p}%
\end{bmatrix}%
.  \label{configuration_matrix_04}
\end{equation}%
The $a_{i},b_{i}$ here can be either non-negative or negative.

What we want is the data of certain involutions and their fixed point locus.
As a first step, we discuss the information about the action of these
involutions. Each involution generates a $\mathbb{Z}/2\mathbb{Z~}$subgroup
of $G_{0}=\mathrm{Aut}(X)$. Moreover, if we have several involutions which
are pairwise commutative, then they generate a subgroup of $G_{0}$,
isomorphic to a product of several $\mathbb{Z}/2\mathbb{Z}$ groups. As a
subgroup of $G_{0}$, we denote the product of these $\mathbb{Z}/2\mathbb{Z}$
groups by $G$. We have the involution%
\begin{equation}
\iota :X\longrightarrow X,~~~\iota \in G.
\end{equation}%
It acts as follows: We consider the involution $\iota $ which can be written
in the form $\iota =((e_{0},e_{1},e_{2},\cdots ,e_{m})$,$(f_{0},f_{1},f_{2},%
\cdots ,f_{n}))$, with the fixed point locus being curves. If the component $%
e_{i}$ is 1, it changes the sign of its corresponding component in $\mathbb{P%
}^{m}$. If $e_{i}$ is 0, it keeps the corresponding component unchanged. The
$f_{j}$'s are treated in the same manner. For instance, the element $%
((0,0,\cdots ,0)$,$(1,1,\cdots ,0))$ sends ($[z_{0},z_{1},\cdots ,z_{m}]$,$%
[y_{0},y_{1},\cdots ,y_{n}]$) to ($[z_{0},z_{1},\cdots ,z_{m}]$,$%
[-y_{0},-y_{1},\cdots ,y_{n}]$). These involutions preserve the holomorphic
form of $X$. There are two possibilities that a point is fixed under the
action of $G$: (i) Every component got acted on is 0; (ii) Every component
that is unchanged is 0. The fixed point locus of the involution is the
disjoint union of two curves. This observation can be deduced as follows
(this argument also applies to more general group actions): If case (i)
holds, they remain unchanged under the involution. If not, suppose there is
an element $\iota ~$in $G$ which acts on $(y_{0},y_{1})$ non-trivially and
at least one of them does not vanish, then $([z_{0},\cdots
,z_{m}],[y_{0},y_{1},\cdots ,y_{n}])$ is sent to $([z_{0},\cdots
,z_{m}],[-y_{0},-y_{1},\cdots ,y_{n}])$ by this element. By the definition
of projective spaces, these two points can coincide only when one of them is
the other one multiplied by $-1$. In this case, those components that remain
unchanged have to be 0, which is case (ii). These are curves inside $X$. The
first component $\Gamma _{1}$ of the curves is defined by $y_{0}=y_{1}=0$,
and this curve can be viewed as a curve in $\mathbb{P}^{m}\times \mathbb{P}%
^{n-2}$ which has coordinates $([z_{0},z_{1},\cdots ,z_{m}],[y_{2},\cdots
,y_{n}])$. The second component $\Gamma _{2}$ of the curves is defined by $%
y_{2}=y_{3}=\cdots =y_{n}=0$, and this curve can be viewed as a curve in $%
\mathbb{P}^{m}\times \mathbb{P}^{1}$ which has coordinates $%
([z_{0},z_{1},\cdots ,z_{m}],[y_{0},y_{1}])$. Due to that the condition $%
y_{0}=y_{1}=y_{2}=\cdots =y_{n}=0$ is not possible in $\mathbb{P}^{n}$,
these two components will not intersect. The fixed point locus of the
involution is%
\begin{equation}
Z=\Gamma _{1}\cup \Gamma _{2}.  \label{curves_locus}
\end{equation}%
It is a disjoint union, since condition (i) and (ii) do not occur at the
same time.

\subsection{Genus Formula of Curves as Generalized Complete Intersections in
Product of Projective Spaces}

\label{sec_genus_formula_subsection}

In this section, we derive a general genus formula for generalized complete
intersection curves in product of projective spaces. The curve itself is a
generalized complete intersection manifold, in that the configuration matrix
for the curve allows negative integer matrix elements. $A$ is the ambient
space, which is the product of projective spaces. $\Gamma $ will denote the
curve. The entries of the configuration matrix of curves in this section
allow negative integers.

Denote the configuration matrix of this family of curves by
\begin{equation}
\Gamma =%
\begin{matrix}
\mathbb{P}^{m} \\
\mathbb{P}^{n}%
\end{matrix}%
\begin{bmatrix}
a_{1} &  & a_{2} &  & a_{3} &  & \cdots &  & a_{N} \\
b_{1} &  & b_{2} &  & b_{3} &  & \cdots &  & b_{N}%
\end{bmatrix}%
\end{equation}%
where $N=m+n-1$, with $m,n\in \mathbb{Z}_{>0}$, and the matrix elements $%
a_{i},b_{i}$ here can be either non-negative or negative integers.

\begin{thm}
The genus $g$ of the above family of curves $\Gamma $ is given by:%
\begin{equation}
\begin{array}{lr}
g=\binom{(\sum_{i=1}^{m+n-1}a_{i})-1}{(\sum_{i=1}^{m+n-1}a_{i})-m-1}\binom{%
(\sum_{i=1}^{m+n-1}b_{i})-1}{(\sum_{i=1}^{m+n-1}b_{i})-n-1}+ &  \\
(\sum_{i_{1}<i_{2}<\cdots <i_{m-1}}\binom{(-\sum_{j=1}^{m-1}a_{i_{j}})-1}{%
(-\sum_{j=1}^{m-1}a_{i_{j}})-m-1})(\sum_{i_{1}<i_{2}<\cdots <i_{m-1}}\binom{%
(-\sum_{j=1}^{m-1}b_{i_{j}})+n}{n})+ &  \\
(\sum_{i_{1}<i_{2}<\cdots <i_{n-1}}\binom{(-\sum_{j=1}^{n-1}a_{i_{j}})+m}{m}%
)(\sum_{i_{1}<i_{2}<\cdots <i_{n-1}}\binom{(-\sum_{j=1}^{n-1}b_{i_{j}})-1}{%
(-\sum_{j=1}^{n-1}b_{i_{j}})-n-1}). &
\end{array}
\label{genus_formula}
\end{equation}%
Here the brackets $(%
\begin{array}{c}
\cdot \\
\cdot%
\end{array}%
)~$mean the binomial coefficients.
\end{thm}

\textit{Proof.} We will use Koszul complex and spectral sequence to derive
the result. The Koszul complex, where the last term is the $m+n-1$-wedge,
is:
\begin{equation}
0\longrightarrow \wedge ^{n+m-1}\mathcal{E}^{\ast }\longrightarrow \cdots
\longrightarrow \wedge ^{2}\mathcal{E}^{\ast }\longrightarrow \mathcal{E}%
^{\ast }\longrightarrow O_{A}\longrightarrow O_{\Gamma }.
\end{equation}

We consider the double complex formed by the cohomology groups of these
sheaves. The ideal sheaf is $\mathcal{E}^{\ast }=\oplus
_{i}O_{A}(-a_{i},-b_{i})$. Moreover, $\wedge ^{k}\mathcal{E}^{\ast }$ are
the anti-symmetric operations on $(\mathcal{E}^{\ast })^{k}$, where the
power stands for tensor product. So we have $\wedge ^{k}\mathcal{E}^{\ast
}=\oplus _{i_{1}<i_{2}<\cdots <i_{k}}\otimes
_{j}O_{A}(-a_{i_{j}},-b_{i_{j}})=\oplus _{i_{1}<i_{2}<\cdots
<i_{k}}O_{A}(-\sum_{j}a_{i_{j}},-\sum_{j}b_{i_{j}})$.

By K\"{u}nneth formula, we expand%
\begin{equation}
H^{q}(A,O_{A}(-\sum_{j}a_{i_{j}},-\sum_{j}b_{i_{j}}))=\sum_{\alpha +\beta
=q}H^{\alpha }(\mathbb{P}^{m},O_{\mathbb{P}^{m}}(-\sum_{j}a_{i_{j}}))\otimes
H^{\beta }(\mathbb{P}^{n},O_{\mathbb{P}^{n}}(-\sum_{j}a_{i_{j}})).
\end{equation}%
Recall the Bott formula for computation of these cohomology. The only
possibly nontrivial terms would be the 0-th or $m$-th cohomology groups for $%
\mathbb{P}^{m}$, and the 0-th or $n$-th cohomology groups for $\mathbb{P}%
^{n} $, depending on the sum of degrees to be positive or not.

Since in the above spectral sequence only $H^{k}(A,\wedge ^{k-1}\mathcal{E}%
^{\ast })$ are mapped to $H^{1}(\Gamma ,O_{\Gamma })$, the terms we need to
consider are as follows:%
\begin{equation}
T_{1}:=\oplus _{i_{1}<i_{2}<\cdots <i_{m-1}}H^{m}(\mathbb{P}^{m},O_{\mathbb{P%
}^{m}}(-\sum_{j=1}^{m-1}a_{i_{j}}))\otimes H^{0}(\mathbb{P}^{n},O_{\mathbb{P}%
^{n}}(-\sum_{j=1}^{m-1}b_{i_{j}})).
\end{equation}%
\begin{equation}
T_{2}:=\oplus _{i_{1}<i_{2}<\cdots <i_{n-1}}H^{0}(\mathbb{P}^{m},O_{\mathbb{P%
}^{m}}(-\sum_{j=1}^{n-1}a_{i_{j}}))\otimes H^{n}(\mathbb{P}^{n},O_{\mathbb{P}%
^{n}}(-\sum_{j=1}^{n-1}b_{i_{j}})).
\end{equation}%
\begin{equation}
T_{3}:=H^{m}(\mathbb{P}^{m},O_{\mathbb{P}^{m}}(-\sum_{i=1}^{m+n-1}a_{i}))%
\otimes H^{n}(\mathbb{P}^{n},O_{\mathbb{P}^{n}}(-\sum_{i=1}^{m+n-1}b_{i})).
\end{equation}

Next we consider $T_{1}$, and $T_{2}$ can be treated in the same manner.
There are restrictions on the indices. The $0$-th cohomology groups would be
non-zero only when $-\sum_{j=1}^{m-1}b_{i_{j}}\geq 0$. They have dimensions $%
\binom{(-\sum_{j=1}^{m-1}b_{i_{j}})+n}{n}$ in this situation. Here the
brackets $\binom{\alpha }{\beta }$ mean the binomial coefficients $\frac{%
\alpha !}{\beta !(\alpha -\beta )!}$. We denote the set of such $m-1$%
-indices $(i_{1},i_{2},\cdots ,i_{m-1})$ as $I_{1}$.

As for the $m$-th cohomology groups, they will be non-zero only when $%
-\sum_{j=1}^{m-1}a_{i_{j}}\leq -(m+1)$, being of dimensions $\binom{%
(-\sum_{j=1}^{m-1}a_{i_{j}})-1}{(-\sum_{j=1}^{m-1}a_{i_{j}})-m-1}$.

We add all those terms with non-vanishing cohomology groups. We set $\binom{%
\alpha }{\beta }$ to be 0 as long as $\alpha <\beta $ or $\beta <0$. We need
to consider all possible tensor products. So the total dimension of $T_{1}$
is:
\begin{equation}
\mathrm{\dim }T_{1}=(\sum_{i_{1}<i_{2}<\cdots <i_{m-1}}\binom{%
(-\sum_{j=1}^{m-1}a_{i_{j}})-1}{(-\sum_{j=1}^{m-1}a_{i_{j}})-m-1}%
)(\sum_{i_{1}<i_{2}<\cdots <i_{m-1}}\binom{(-\sum_{j=1}^{m-1}b_{i_{j}})+n}{n}%
).
\end{equation}%
In the above, the summation is over the index set $I_{1}$.

We denote $I_{2}$ the set of $n-1$-indices $(i_{1},i_{2},\cdots ,i_{n-1})$
such that $-\sum_{j=1}^{n-1}a_{i_{j}}\geq 0$. We have
\begin{equation}
\mathrm{\dim }T_{2}=(\sum_{i_{1}<i_{2}<\cdots <i_{n-1}}\binom{%
(-\sum_{j=1}^{n-1}a_{i_{j}})+m}{m})(\sum_{i_{1}<i_{2}<\cdots <i_{n-1}}\binom{%
(-\sum_{j=1}^{n-1}b_{i_{j}})-1}{(-\sum_{j=1}^{n-1}b_{i_{j}})-n-1}).
\end{equation}%
Here, the summation is over the index set $I_{2}$.

The only left term is $T_{3}$, which has only one component. By Bott
formula, and since $-\sum_{i=1}^{m+n-1}a_{i}\leq
-(m+1),-\sum_{i=1}^{m+n-1}b_{i}\leq -(n+1)$, we find that its dimension is
\begin{equation}
\mathrm{\dim }T_{3}=\binom{(\sum_{i=1}^{m+n-1}a_{i})-1}{%
(\sum_{i=1}^{m+n-1}a_{i})-m-1}\binom{(\sum_{i=1}^{m+n-1}b_{i})-1}{%
(\sum_{i=1}^{m+n-1}b_{i})-n-1}.
\end{equation}

{\hfill $\square $}

This is a general formula for the genus of smooth curves as generalized
complete intersections, as a function of the data of the configuration
matrix. This formula is useful when we calculate the Hodge numbers of the
blow up of the generalized complete intersection manifolds along curves.

We also mention a special case as follows. For the simplest example of the
product space of projective spaces, we consider the special case $A=\mathbb{P%
}^{1}\times \mathbb{P}^{1}$ with coordinates $[z_{0},z_{1}],[y_{0},y_{1}]$
respectively. For example, a polynomial $F$ of bi-degree $(d_{1},d_{2})$
defines a curve $C$. Below, we will also add singular points $P$ on the
Riemann surface, for a more generality. We have the following:

\begin{prop}
\label{prop_genus} For $C$ as above, its genus g is given by:
\begin{equation}
g=(d_{1}-1)(d_{2}-1)-\sum_{P}\frac{r_{P}(r_{P}-1)}{2}.  \label{eqn_g_04}
\end{equation}%
where $r_{P}$ is the multiplicity of $P$ in $C$.
\end{prop}

The smooth part of this formula (\ref{eqn_g_04}) is a special case of our
above general formula (\ref{genus_formula}), when we plug in $m=1$, $n=1$, $%
a_{1}=d_{1}~$and $b_{1}=d_{2}$. We prove the genus formula in this form in
Appendix A. Among other things, the main tool we are going to use is the
Riemann-Roch theorem.

\subsection{Hodge Numbers of Blow Up}

\label{sec_blow_up_subsection}

We have considered involutions of gCICY, and then identified some fixed
point locus of the involutions. We can then also perform a blow-up on the
manifold. Since the involutions preserve the holomorphic form, the blow up
along the fixed point locus, is still Calabi-Yau. Finally, we can derive the
Hodge numbers of the blow ups from the cohomology of the exceptional
divisors, for example \cite{Voisin}. Moreover, we relate the genus of the
curves to the Hodge numbers. In conjunction with the genus formula (\ref%
{genus_formula}) in section \ref{sec_genus_formula_subsection}, we will
determine the Hodge numbers of the blow up of the manifold. We consider a
special case of threefolds (\ref{configuration_matrix_04}), whose
configuration matrix is
\begin{equation}
X=%
\begin{matrix}
\mathbb{P}^{1} \\
\mathbb{P}^{4}%
\end{matrix}%
\begin{bmatrix}
a_{1} &  & a_{2} \\
b_{1} &  & b_{2}%
\end{bmatrix}%
,  \label{example_03}
\end{equation}%
where $a_{i}$ and $b_{i}$ can be either non-negative or negative integers,
and the Calabi-Yau condition is $a_{2}=2-a_{1}$ and $b_{2}=5-b_{1}$. The
range of the matrix elements in (\ref{example_03}) has been classified in
\cite{Anderson:2015iia,Berglund:2016yqo}. When $b_{1}=1$, then $a_{1}\geq 0$%
. When $b_{1}=2$, then $0\leq a_{1}\leq 3$, in which case $a_{1}$ has an
upper bound due to the requirement of the existence of a section of $%
O(2-a_{1},3)$ over the subvariety defined by the first column \cite%
{Berglund:2016yqo}. We denote the coordinates of $\mathbb{P}^{1}~$and$~%
\mathbb{P}^{4}$ by $[z_{0},z_{1}],[y_{0},y_{1},y_{2},y_{3},y_{4}]$
respectively. The group that we will consider to act on $A=\mathbb{P}%
^{1}\times \mathbb{P}^{4}$ would be $G=(\mathbb{Z}/2\mathbb{Z})\times (%
\mathbb{Z}/2\mathbb{Z})$. In $G$, the action ((0,0),(1,1,0,0,0)) represents
the involution:
\begin{equation}
\iota :X\longrightarrow
X.~~([z_{0},z_{1}],[y_{0},y_{1},y_{2},y_{3},y_{4}])\mapsto
([z_{0},z_{1}],[-y_{0},-y_{1},y_{2},y_{3},y_{4}]).
\end{equation}

By the consideration in Section \ref{sec_fixed_point_locus_subsection}, $%
\iota $ fixes a point if one of these two situations happens: $y_{0}=y_{1}=0$
or $y_{2}=y_{3}=y_{4}=0$. The second situation arises since in the
projective space, we can multiply all coordinates by $-1$ without changing
the point. Denote these two parts of fixed locus by $\Gamma _{1},\Gamma _{2}$
respectively.

Now we consider the example with $a_{1}=3$, $b_{1}=2$, which is the
threefold considered in \cite{Anderson:2015iia,Garbagnati:2017rtb}, which
gives $X=%
\begin{matrix}
\mathbb{P}^{1} \\
\mathbb{P}^{4}%
\end{matrix}%
\begin{bmatrix}
3 &  & -1 \\
2 &  & 3%
\end{bmatrix}%
$. On $\Gamma _{1}$, we have $y_{0}=y_{1}=0$, which reduces $A$ to be $%
\tilde{A}_{1}=\mathbb{P}^{1}\times \mathbb{P}^{2}$, whose coordinates are $%
([z_{0},z_{1}],[y_{2},y_{3},y_{4}])$. In this case, $\Gamma _{1}$ would be
\begin{equation}
\Gamma _{1}=%
\begin{matrix}
\mathbb{P}^{1} \\
\mathbb{P}^{2}%
\end{matrix}%
\begin{bmatrix}
3 &  & -1 \\
2 &  & 3{}%
\end{bmatrix}%
.
\end{equation}%
This curve is a generalized complete intersection manifold in $\mathbb{P}%
^{1}\times \mathbb{P}^{2}$. Using our general formula (\ref{genus_formula})
in Section \ref{sec_genus_formula_subsection} yields that this curve has
genus $g(\Gamma _{1})=8$.

Next we consider $\Gamma _{2}$. On $\Gamma _{2}$, we would have $%
y_{2}=y_{3}=y_{4}=0$, and this reduces $A$ to be $\tilde{A}_{2}=\mathbb{P}%
^{1}\times \mathbb{P}^{1}$ which has coordinates $%
([z_{0},z_{1}],[y_{0},y_{1}])$. Consequently, this part of fixed locus is a
curve in $\mathbb{P}^{1}\times \mathbb{P}^{1}$, which is%
\begin{equation}
\Gamma _{2}=%
\begin{matrix}
\mathbb{P}^{1} \\
\mathbb{P}^{1}%
\end{matrix}%
\begin{bmatrix}
3 \\
2%
\end{bmatrix}%
.
\end{equation}%
Using our general formula (\ref{genus_formula}) in Section \ref%
{sec_genus_formula_subsection} yields that its genus is $g(\Gamma _{2})=2$.

The genus gives the data of Hodge numbers. We have that $\mathrm{\dim }%
H^{1}(\Gamma _{j},O_{\Gamma _{j}})=g(\Gamma _{j})$, $j=1,2$, and hence $%
\mathrm{\dim }H^{1}(\Gamma _{j},\mathbb{Z})=2g(\Gamma _{j})$~and $\mathrm{%
\dim }H^{0}(\Gamma _{j},\mathbb{Z})=1$. Now we make a blow-up of the
original gCICY along the curves, that is in the form (\ref{curves_locus}).
The involution preserves the holomorphic form, and hence the blow-up along
the fixed point locus of the involution is still a Calabi-Yau.

The Hodge structure of the blow up is described by the following theorem
\cite{Voisin}.

\begin{thm}
Denote $X_{Z}$ to be the blow up of $X$ along $Z$. We have the following
isomorphism of Hodge structures:
\begin{equation}
H^{k}(X,\mathbb{Z})\oplus (\oplus _{i=0}^{r-2}H^{k-2i-2}(Z,\mathbb{Z}%
))\simeq H^{k}(X_{Z},\mathbb{Z}),
\end{equation}%
where $r$ is the codimension of $Z$ in $X$, and $r-1=rank(E)$, with $E:=\tau
^{-1}(Z)$ being the exceptional divisor viewed as a bundle over $Z$.
\end{thm}

Here, the exceptional divisor $E:=\tau ^{-1}(Z)$ is a $\mathbb{P}^{1}$
bundle over the curves. When we take $k=0,1$, we know that $h^{0}(X)$ and$%
~h^{1}(X)$ would stay unchanged when we take the blowing-up. The change
happens in the case where $k=2,3$. The only non-trivial summand involving $Z$
would be those ones corresponding to $i=0$. Consider the case $k=2$. The
term from $Z$ is $H^{0}(Z,\mathbb{Z})$. These two components will not
intersect since the coordinates of them can not all be zero. Hence every
component will increase $h^{2}(X_{Z})$ by 1. Hence totally it will increase
by $\dim H^{0}(Z,\mathbb{Z})=2$.~We have $h^{2}(X_{Z})=h^{2}(X)+2=4~$and$%
~h^{1,1}(X_{Z})=4$.$~$In the case $k=3$, the summand involving $Z$ would be $%
H^{1}(Z,\mathbb{Z})$, offering additional dimension $\dim H^{1}(Z,\mathbb{Z}%
)=2g(\Gamma _{1})+2g(\Gamma _{2})$,~which, in this example, is $20$. Hence $%
h^{3}(X_{Z})=h^{3}(X)+20=94+20=114$. Since CY condition gives us $%
h^{3,0}(X_{Z})=1$, we know that $h^{2,1}(X_{Z})=\frac{1}{2}(114-2)=56$, in
this case.

Our genus formula of the curves (\ref{genus_formula}) is needed when we
consider Hodge numbers of blow-up of gCICYs. These blow-ups can provide new
Calabi-Yau manifolds and their variants in the moduli space of Calabi-Yaus.
They are also useful for string compactification and non-perturbative
superpotentials in lower dimensional field theory after the compactification.


\section{Spectral Sequence Approach}

\label{sec_spectral_sequence_approach}

\renewcommand{\theequation}{6.\arabic{equation}} \setcounter{equation}{0} %
\renewcommand{\thethm}{6.\arabic{thm}} \setcounter{thm}{0}

The previous method requires much effort in computation, whereas the
spectral sequence approach needs much less elaboration since it can treat
all sheaves in the Koszul sequences of four or more terms at one time. This
is related to the codimension of the submanifolds as follows: the Koszul
sequence of a submanifold of codimension $d$ is of length $d+2$. So if the
codimension is bigger, we need to treat Koszul sequences of more terms.

\subsection{Spectral Sequences}

When the sequence is no longer of three terms, spectral sequence truly
demonstrates its power comparing with iterative usage of short exact
sequences. We denote $O_{X}(b_{1},\cdots ,b_{t}):=O_{A}(b_{1},\cdots
,b_{t})|_{X}$, and we use spectral sequences to compute%
\begin{equation}
h^{\ast }(X,O_{X}(b_{1},\cdots ,b_{t})).
\end{equation}%
We use $h^{\ast }(X,F)$ to denote the list of dimensions of cohomology
groups ($h^{0},h^{1},h^{2},\cdots $) where $h^{i}(X,F)=\dim H^{i}(X,F)$.

As we have explained, the manifold $X$ can be construced by line bundles.
Suppose we have line bundles $E_{i}$ respectively. The number of columns in
the configuration matrix is $K$, and hence there are $K$ line bundles. Then
when we consider the restriction map $r:O_{A}\longrightarrow O_{X}$, the
kernel would be the dual of $E:=\oplus _{i=1}^{K}E_{i}$. And the map from $%
E^{\ast }$ to $O_{A}$ can be obtained by contracting with the dual $E$.
Analyzing kernels of these maps iteratively, we find a sequence:
\begin{equation}
0\longrightarrow \wedge ^{K}E^{\ast }\longrightarrow \cdots \longrightarrow
\wedge ^{2}E^{\ast }\longrightarrow E^{\ast }\longrightarrow
O_{A}\longrightarrow O_{X}\longrightarrow 0.  \label{10a}
\end{equation}%
For brevity, in the above notation, we have used the brief notation $O_{X}$
to mean $i_{\ast }O_{X}$, where $i:X\rightarrow A~$is the inclusion. When we
want to derive data on cohomology groups of other sheaves, we can tensor
this sequence by locally free sheaves or bundles on $A$.

For example, in our case in Section \ref{sec_blow_up_subsection}, $%
E=E_{1}\oplus E_{2}$. For instance, for $a_{1}=3$, $b_{1}=2$, we have $%
E_{1}=O_{A}(3,2),\,E_{2}=O_{A}(-1,3)$. Hence $E^{\ast }=O_{A}(-3,-2)\oplus
O_{A}(1,-3),\,\wedge ^{2}E^{\ast }=O_{A}(-2,-5)$. We notice that $\wedge
^{2}E^{\ast }=O_{A}(-2,-5)=K_{A}$ is the canonical bundle of $A$. This is
not a coincidence. As a matter of fact, this is a criterion of Calabi-Yau
condition.

The compact manifold $X$ is a Calabi-Yau threefold if the structure sheaf $%
O_{X}$ fits into a resolution sequence:
\begin{equation}
0\longrightarrow F_{K}^{\ast }\longrightarrow \cdots \longrightarrow
F_{2}^{\ast }\longrightarrow F_{1}^{\ast }\longrightarrow F_{0}^{\ast
}\longrightarrow O_{X}\longrightarrow 0.
\end{equation}%
Here, we require $F_{0}^{\ast }=O_{A},\,F_{K}^{\ast }=K_{A}$, the canonical
bundle of $A$. We can tensor the above sequence with a locally free sheaf $F$
of interest, and make an double complex $W^{p,q}$ of their sheaf comhomology
$H^{q}(A,F_{p}^{\ast }\otimes F)$.

We will then turn to the example we have considered in Section \ref%
{sec_blow_up_subsection}. The goal here is to use the powerful tool spectral
sequence to determine Hodge numbers. Our approach using the spectral
sequence of double complex is an improvement of the approach using multiple
short exact sequences.

For the example in Section \ref{sec_blow_up_subsection}, the sequence (\ref%
{10a}) becomes%
\begin{equation}
0\longrightarrow \wedge ^{2}E^{\ast }\longrightarrow E^{\ast
}\longrightarrow O_{A}\longrightarrow O_{X}\longrightarrow 0.
\end{equation}%
More specifically, in this example, the above sequence is
\begin{equation}
0\longrightarrow O_{A}(-2,-5)\longrightarrow O_{A}(-3,-2)\oplus
O_{A}(1,-3)\longrightarrow O_{A}\longrightarrow O_{X}\longrightarrow 0.
\end{equation}%
To obtain $h^{\ast }(X,O_{X}(-1,3))$, we tensor this sequence by the sheaf $%
F=O_{A}(-1,3)$:
\begin{equation}
0\longrightarrow O_{A}(-3,-2)\longrightarrow O_{A}(-4,1)\oplus
O_{A}\longrightarrow O_{A}(-1,3)\longrightarrow O_{X}(-1,3)\longrightarrow 0.
\end{equation}

Other ingredients we need would be the cohomology groups of sheaves involved
in this sequence. We build the following double complex $W^{p,q}$ consisting
of their cohomology groups:%
\begin{equation}
\xymatrix{ H^{0}(A,O_{A}(-3,-2)) \ar[r] \ar[d] & H^{0}(A,O_{A}(-4,1)\oplus
O_{A})\ar[r] \ar[d] & H^{0}(A,O_{A}(-1,3)) \ar[r] \ar[d] &
H^{0}(X,O_{X}(-1,3)) \ar[d] \\ H^{1}(A,O_{A}(-3,-2)) \ar[r] \ar[d] &
H^{1}(A,O_{A}(-4,1)\oplus O_{A}) \ar[r] \ar[d] & H^{1}(A,O_{A}(-1,3)) \ar[r]
\ar[d] & H^{1}(X,O_{X}(-1,3)) \ar[d] \\ \vdots & \vdots & \vdots & \vdots }
\end{equation}%
In the double complex above, the vertical direction is labeled by $q$, and
the horizontal direction is labeled by $p$. We have used the fact that $%
H^{i}(X,O_{X}(b_{1},\cdots ,b_{t}))=H^{i}(A,i_{\ast }O_{X}(b_{1},\cdots
,b_{t}))$, for each $i$, which can be easily derived from Lemma 2.10 of Ch.
III in \cite{Hartshorne}.

Those cohomology groups involving $A$ can be computed through Bott formula
and K\"{u}nneth formula. We have that
\begin{equation}
h^{\ast }(A,O_{A}(-1,3))=(0,0,0,0,0,0).  \label{101}
\end{equation}%
Similarly,
\begin{equation}
h^{\ast }(A,O_{A}(-3,-2))=(0,0,0,0,0,0).  \label{102}
\end{equation}%
The next sheaf we consider is $O_{A}(-4,1)$:
\begin{equation}
h^{\ast }(A,O_{A}(-4,1))=(0,15,0,0,0,0).  \label{103}
\end{equation}%
And finally, a familiar one:
\begin{equation}
h^{\ast }(A,O_{A})=(1,0,0,0,0,0).  \label{104}
\end{equation}

The dimensions of the cohomology groups of sheaves in the double complex $%
W^{p,q}$ are in the following table:
\begin{equation}
\begin{matrix}
O_{A}(-3,-2) &  & O_{A}(-4,1)\oplus O_{A} &  & O_{A}(-1,3) &  & O_{X}(-1,3)
&  &  \\
0 &  & 1 &  & 0 &  & h^{0}(X,O_{X}(-1,3))=14 &  &  \\
0 &  & 15 &  & 0 &  & h^{1}(X,O_{X}(-1,3))=0 &  &  \\
\vdots &  & \vdots &  & \vdots &  & \vdots &  &  \\
&  &  &  &  &  &  &  &
\end{matrix}%
\end{equation}

In the double complex context, since the $r$-th differential operators of a
spectral sequence would send an element $r$ steps rightward and $r-1$ steps
upward, the only $H^{k}(X,O_{X}(-1,3))$ that possibly has a source would be $%
H^{0}(X,O_{X}(-1,3))$. Its dimension can be calculated from counting the
Euler characteristic. The Euler characteristic is counted with respect to
the total complex $M^{k}:=\oplus _{p+q=k}W^{p,q}$. The Euler characteristic
number of the complex of sheaves equals the Euler characteristic number of
the total complex $M^{k}$. This is used when we calculate the dimensions of
the abelian groups involved. We have then%
\begin{equation}
h^{\ast }(X,O_{X}(-1,3))=(14,0,0,0).
\end{equation}

\subsection{Genus of Curves}

We can also use spectral sequences of double complexes to compute the genus
of curves. Recall that when we want to determine the Hodge numbers of the
blow up of $X$, we have determined the genus of the curves along which it is
blown-up. The genus of them can be determined by our genus formula (\ref%
{genus_formula}) in Section \ref{sec_genus_formula_subsection}. For example,
one of the curves is characterized as:
\begin{equation}
\Gamma _{1}=%
\begin{matrix}
\mathbb{P}^{1} \\
\mathbb{P}^{2}%
\end{matrix}%
\begin{bmatrix}
3 &  & -1 \\
2 &  & 3%
\end{bmatrix}%
.
\end{equation}%
We can also compute similarly for the other curve $\Gamma _{2}$. By the
relation $g(\Gamma _{1})=\dim H^{1}(\Gamma _{1},O_{\Gamma _{1}})$, we turn
this into a sheaf theory computation. Now we take $\tilde{A}=\mathbb{P}%
^{1}\times \mathbb{P}^{2}$. We have a similar four-term sequence:
\begin{equation}
0\longrightarrow O_{\tilde{A}}(-2,-5)\longrightarrow O_{\tilde{A}%
}(-3,-2)\oplus O_{\tilde{A}}(1,-3)\longrightarrow O_{\tilde{A}%
}\longrightarrow O_{\Gamma _{1}}\longrightarrow 0.
\end{equation}%
We build the following double complex $W^{p,q}$ consisting of their
cohomology groups:
\begin{equation}
\xymatrix{ H^{0}(\tilde{A},O_{\tilde{A}}(-2,-5)) \ar[r] \ar[d] &
H^{0}(\tilde{A},O_{\tilde{A}}(-3,-2)\oplus O_{\tilde{A}}(1,-3)) \ar[r]
\ar[d] & H^{0}(\tilde{A},O_{\tilde{A}}) \ar[r] \ar[d] & H^{0}(\Gamma
_{1},O_{\Gamma _{1}}) \ar[d] \\ H^{1}(\tilde{A},O_{\tilde{A}}(-2,-5))\ar[r]
\ar[d] & H^{1}(\tilde{A},O_{\tilde{A}}(-3,-2)\oplus
O_{\tilde{A}}(1,-3))\ar[r] \ar[d] & H^{1}(\tilde{A},O_{\tilde{A}})\ar[r]
\ar[d] & H^{1}(\Gamma _{1},O_{\Gamma _{1}}) \ar[d] \\ \vdots & \vdots &
\vdots & \vdots }
\end{equation}

The dimensions of the cohomology groups we need are listed below:
\begin{equation}
\begin{array}{lr}
h^{\ast }(\tilde{A},O_{\tilde{A}}(-2,-5))=(0,0,0,6), &  \\
h^{\ast }(\tilde{A},O_{\tilde{A}}(-3,-2))=(0,0,0,0), &  \\
h^{\ast }(\tilde{A},O_{\tilde{A}}(1,-3))=(0,0,2,0), &  \\
h^{\ast }(\tilde{A},O_{\tilde{A}})=(1,0,0,0). &
\end{array}%
\end{equation}

The dimensions of their cohomology groups in the double complex $W^{p,q}$
are:
\begin{equation}
\begin{matrix}
O_{\tilde{A}}(-2,-5) &  & O_{\tilde{A}}(-3,-2)\oplus O_{\tilde{A}}(1,-3) &
& O_{\tilde{A}} &  & O_{\Gamma _{1}} \\
0 &  & 0 &  & 1 &  & h^{0}(\Gamma _{1},O_{\Gamma _{1}})=1 \\
0 &  & 0 &  & 0 &  & h^{1}(\Gamma _{1},O_{\Gamma _{1}})=8 \\
0 &  & 2 &  & 0 &  & h^{2}(\Gamma _{1},O_{\Gamma _{1}})=0 \\
6 &  & 0 &  & 0 &  & 0 \\
\vdots &  & \vdots &  & \vdots &  & \vdots%
\end{matrix}%
\end{equation}

In the table above, the three-level differential operator will map the
cohomology group where the 6 seats, to $H^{1}(\Gamma _{1},O_{\Gamma _{1}})$,
and the two-level operator would send the 2 to $H^{1}(\Gamma _{1},O_{\Gamma
_{1}})$. Except these two places, there are no other cohomology groups that
can be mapped to any one of them, forming images to reduce their dimensions
when we take cohomology. And, lower level operators would map it to the
cohomology groups where the 0 seats, which means that they would keep to be
kernels, remaining unchanged when we take cohomology. Hence they would be
stable until they touch $H^{1}(\Gamma _{1},O_{\Gamma _{1}})$. We can also
count the Euler characteristic as before. In conclusion, we have $%
g=h^{1}(\Gamma _{1},O_{\Gamma _{1}})=2+6=8$, which is exactly the genus of
this curve. This agrees with our general genus formula (\ref{genus_formula})
in Section \ref{sec_genus_formula_subsection}.

\section{Discussion}

\label{sec_discussion}

We worked out aspects of cohomology of sheaves on the generalized complete
Calabi-Yau manifolds, and developed some tools and approaches to understand
them better. These manifolds are constructed from line bundles in ambient
product spaces. For the generalized case, one can make use of line bundles
which do not have global sections on the ambient space, but have sections
when restricted to appropriate subvarieties. One then constructs the
generalized complete intersection Calabi-Yau in these subvarieties. Hence
approaches and methods of sheaves are very necessary for these generalized
constructions. The main tools we used include cohomology of sheaves, Bott
formula, K\"{u}nneth formula, and spectral sequences.

\vspace{1pt}

We have presented and proved a vanishing theorem of cohomology groups of
sheaves which naturally generalized the original one for a single projective
space to the case for a product of several projective spaces. The technique
we used is to twist the Poincare residue short exact sequences. Next we
considered the long exact sequences of cohomology groups associated with
them and proved the vanishing theorem. This is useful for computing the
Hodge numbers of complete intersection and generalized complete intersection
manifolds in a product of projective spaces.

\vspace{1pt}

In the version of the vanishing theorem for a single projective space, the
vanishing is related to the fact that the~action of the projective linear
group $PGL_{n+2}(k)$ on degree $d$ hypersurface in $\mathbb{P}^{n+1}$
induces a morphism which is proper and finite. This in turn is useful for
the Haar measure of the matrix integrals for $PGL_{n+2}(k)$ \cite{Katz
Sarnak}. It would be good to understand whether there are generalizations of
these relations in the cases of a product of projective spaces.

\vspace{1pt}

Moreover, we proved an equivalence between configuration matrices for
complete intersection Calabi-Yau manifolds. The manifolds from the
configuration matrices of the two sides are birationally equivalent.
Further, if the manifolds of the both sides have nef canonical bundles and
are smooth, then they also have the same Hodge numbers \cite{Ito}. There are
also other types of equivalences between configuration matrices \cite%
{Anderson:2015iia,Candelas:1987kf,Berglund:2016yqo}. Besides identities and
equivalences, there are other types of relationships between configuration
matrices pertaining to deformations and geometric transitions of the
varieties \cite{Anderson:2015iia,Berglund:2016yqo}. These equivalences and
relationships are important and useful in the classification of gCICY \cite%
{Anderson:2015iia}.

\vspace{1pt}

Further, we have identified some involutions of gCICY. We identified some
fixed point locus of involutions of gCICY, which are curves. We considered
blow-up of gCICY along the curves, which can produce new Calabi-Yau
manifolds. The blow-ups along the fixed point locus of the involutions
provided new Calabi-Yaus and their variants, which can be useful for string
compactification and non-perturbative superpotentials after the
compactification. This enlarged the range of the construction of Calabi-Yau
manifolds, and is important for the moduli space of Calabi-Yau manifolds, in
which topologically distinct Calabi-Yau manifolds may be connected to each
other by geometric transitions \cite{Reid,Gross}, such as via blowing-downs
and blowing-ups. And this process of geometric transitions could also
involve \cite{Lin:2014lya} non-K{\"{a}}hler Calabi-Yau manifolds, which
include complex non-K{\"{a}}hler manifolds with a trivial canonical bundle.

\vspace{1pt}

We presented a genus formula for curves in generalized complete intersection
manifolds, which is useful for computing the Hodge numbers of blow-ups of
the generalized complete intersection manifolds along the curves. These
curves themselves can also be viewed as generalized complete intersection
manifolds. Since these curves are the fixed point locus of involutions which
preserve the holomorphic forms, it would be interesting to identify the
fixed point locus of more general quotient symmetries in the automorphism
groups.

\vspace{1pt}

Many of the construction of gCICYs have K3-fibrations and elliptic
fibrations \cite{Berglund:2016yqo,Berglund:2016nvh,Anderson:2015iia}, which
are widely useful for string dualities, because of the fiber structures,
such as the heterotic/IIA duality and heterotic/F-theory duality. These
fibrations can widely appear in the context of heterotic string theory, for
example \cite%
{Candelas:1985en,Heckman:2013sfa,Melnikov:2014ywa,Lin:2014lya,Lin:2016bzq}
and references therein. In the context of heterotic theory, it would be very
interesting to consider, in addition, vector bundles on these generalized
complete intersection Calabi-Yau manifolds.

\vspace{1pt}

Furthermore, we used a spectral sequence approach that facilitates the
computations of the cohomology group of sheaves of the generalized complete
intersection manifolds. It is particularly useful for subvarieties with
codimension one or higher. This approach is useful for computing the Hodge
numbers and the genus of curves in generalized complete intersection
manifolds. These Hodge numbers play an important role when we consider the
deformation classes.

\vspace{1pt}

Constructing examples of Calabi-Yau manifolds is an interesting theme due to
mirror symmetry and the search for mirror dual manifolds. Some of the
generalized complete intersection manifolds would be new to previous
classifications. Some of the generalized models may be related to weighted
projective spaces by nontrivial blowing-downs \cite{Berglund:2016nvh}. One
can describe mirror dual manifolds by transposition of the degree matrices
\cite{Berglund:1991pp,Batyrev:1994hm,Hosono:1994ax,Berglund:1994qk} in
weighted projective spaces. It may be interesting to find mirror dual
manifolds of these new models and to understand the mirror duality from the
point of view of the worldsheet theory better.

\vspace{1pt}

These new types of Calabi-Yau manifolds open up new possibilities for
analysis on the worldsheet theory of strings on these spaces as target
spaces, and may give new insights in the context of gauged linear sigma
models \cite{Witten:1993yc,Berglund:2016nvh,Sharpe:2015vza}. The defining
equations with negative degrees for the line bundles involve Laurent
polynomials, which have their poles. However, their poles are carefully
avoided through beforehand intersection. The superpotential terms on the
worldsheet would involve Laurent polynomials, which would be a new
phenomenon.

\vspace{1pt}

The CICYs in products of projective spaces have also enabled the computation
of instanton corrections in string theory compactifications, see for example
\cite{Beasley:2005iu,Witten:1996bn}. There are new divisors in gCICY that
are not from the divisors of the ambient product space of projective spaces.
One can wrap branes inside these new divisors in gCICY. Many ordinary CICYs
also have divisors that do not descend from the ambient space, and the ones
where all divisors do descend are `favorable CICYs' \cite{Anderson:2017aux}.
Euclidean branes wrapping nontrivial new divisors contribute to the
non-perturbative superpotentials in lower dimensional field theory after the
compactification, under specific conditions \cite%
{Anderson:2015yzz,Witten:1996bn} on the cohomology groups of sheaves. This
provides new types of instanton corrections in the non-perturbative
superpotentials, when compactified to lower dimensions.

\vspace{1pt}

\section*{Acknowledgments}

We would like to thank X. Gao, S.-J. Lee, B. Wu, and S.-T. Yau for
communications or discussions. The work was supported in part by Yau
Mathematical Sciences Center and Tsinghua University.

\appendix

\section{Another Genus Formula}

\label{appendix_special_case}

\renewcommand{\theequation}{A.\arabic{equation}} \setcounter{equation}{0} %
\renewcommand{\thethm}{A.\arabic{thm}} \setcounter{thm}{0} %
\renewcommand{\theprop}{A.\arabic{prop}} \setcounter{prop}{0}

In this appendix, we prove proposition \ref{prop_genus} in Section \ref%
{sec_genus_formula}. The main tool we are going to use is the Riemann-Roch
theorem. The expression (\ref{eqn_g_04}) is similar to the case of the genus
of curves of a single degree $n$ in $\mathbb{P}^{2}$, which can be denoted
as $\mathbb{P}^{2}[n]$, see for example \cite{Fulton}. Our cases are
different from theirs for the single projective space.

We denote the curves to be $X=%
\begin{matrix}
\mathbb{P}^{1} \\
\mathbb{P}^{1}%
\end{matrix}%
\begin{bmatrix}
d_{1} \\
d_{2}%
\end{bmatrix}%
,$ which are bi-degree $(d_{1},d_{2})$ curves in $A=\mathbb{P}^{1}\times
\mathbb{P}^{1}$. We will pick appropriate lines to intersect $X$ to
construct appropriate divisor for the calculation of the genus. Here we also
introduce a divisor to estimate the irregularity of $X$:
\begin{equation}
E=\sum_{P}(r_{P}-1)P.
\end{equation}%
For a generic point, $r_{P}-1=0$, it would not arise in the above sum.
Another concept we will need is adjointness: A form $G$ is said to be
adjoint to $X$, if and only if $G\geq E$.

We can choose linear forms which are degree-one homogeneous polynomials $%
H_{1},H_{2}$ on the variables of the first and second factor spaces
respectively. The $H_{i}$ will pinpoint the location of the $i$-th factor,
determining the ratio of its coordinates. Then, we are left with an equation
on the variables of the other space factor. Hence a generic choice of $H_{i}$
would generate $d_{j}$ points, $j\neq i$. We denote the equation and its
divisor with the same letter. Set
\begin{equation}
H_{1}\cap X=Q_{1}+\cdots +Q_{d_{2}}:=Q,~~H_{2}\cap X=S_{1}+\cdots
+S_{d_{1}}:=S.
\end{equation}%
We consider%
\begin{equation}
E_{m}=m_{2}S+m_{1}Q-E,
\end{equation}%
where $m=(m_{1},m_{2})$ is a bi-index. For convenience, we introduce $M:=%
\frac{1}{2}\deg \,E=\sum_{P}\frac{r_{P}(r_{P}-1)}{2}$. We have:
\begin{equation}
\deg \,E_{m}=m_{1}d_{2}+m_{2}d_{1}-2M.
\end{equation}

We make the following definitions:%
\begin{equation}
N_{m}:=\{\mathrm{Forms~of~{bi}}\text{\textrm{-}}\mathrm{{\text{d}egree}~}%
m=(m_{1},m_{2})~\mathrm{which~are~adjoint~to~}X\}.
\end{equation}%
\begin{equation}
F_{k}:=\{\mathrm{All\ forms\ of\ \mathrm{bi}}\text{\textrm{\textrm{-}}}%
\mathrm{\mathrm{{\text{d}egree}}~}k=(k_{1},k_{2})\}.
\end{equation}%
We also have that%
\begin{equation}
L(D):=\{f|\mathrm{div}f+D\geq 0\}~~\text{\textrm{and~}}~l(D)=\mathrm{\dim }%
L(D).
\end{equation}

In order to use Riemann-Roch, we need to calculate $l(E_{m})$. To this end,
we will construct an exact sequence involving forms of certain bi-degree
with respect to the homogeneous coordinates.

To calculate $l(E_{m})$, we construct following map:
\begin{equation}
\begin{array}{lr}
\varphi :N_{m}\longrightarrow L(E_{m}), &  \\
\varphi (G)=\frac{G}{H_{1}^{m_{1}}H_{2}^{m_{2}}}. &
\end{array}%
\end{equation}%
Next we consider the kernel of this map, since it sends a form to the
function field of the curve defined by $F=0$, it will be zero if and only if
it is divisible by $F$.

The next result is that $\varphi $ is onto. For any $f=\frac{B}{C}\in
L(E_{m})$ where $B,C$ are forms of the same degree. By definition we have $%
\mathrm{div}(B)-\mathrm{div}(C)+\mathrm{div}(E_{m})\geq 0$. This gives us
forms $\Psi ,\Phi $ such that $BH_{1}^{m_{1}}H_{2}^{m_{2}}=\Psi C+\Phi F$.
After taking restriction to the quotient $F=0$, this gives us $f=\frac{\Psi
}{H_{1}^{m_{1}}H_{2}^{m_{2}}}$. Notice that $\mathrm{div}(\Psi )=\mathrm{div}%
(BH_{1}^{m_{1}}H_{2}^{m_{2}})-\mathrm{div}(C)\geq E$, hence we see that $%
\Psi \in N_{m}$ and $f=\varphi (\Psi )$. Combining above characterization of
$\ker \varphi $ and the surjectivity, we have the following exact sequence:
\begin{equation}
0\longrightarrow F_{m-d}\longrightarrow N_{m}\longrightarrow
L(E_{m})\longrightarrow 0,  \label{g2}
\end{equation}%
where $m=(m_{1},m_{2})$ and $m-d=(m_{1}-d_{1},m_{2}-d_{2})$.

In smooth situation, $F_{k}$ and $N_{k}$ would coincide, both of them have
dimension $(k_{1}+1)(k_{2}+1)$. The exact sequence (\ref{g2}) gives us
\begin{equation}
l(E_{m})=(m_{1}+1)(m_{2}+1)-(m_{1}-d_{1}+1)(m_{2}-d_{2}+1).  \label{g4}
\end{equation}%
Then, using Riemann-Roch we have:
\begin{equation}
g=\deg \,E_{m}-l(E_{m})+1=(d_{1}-1)(d_{2}-1).
\end{equation}

Now we consider a more general case, by adding the correction term of
singular points $P$. To this end, we introduce the following notation: $%
N=N(k;r_{1}J_{1},r_{2}J_{2},\cdots ,r_{a}J_{a})$ is the space of forms of
bi-degree $k=(k_{1},k_{2})$, whose corresponding curves have multipicity at
least $r_{i}$ at point $J_{i}$. Next we will count the dimension of $%
N(m;r_{1}J_{1},r_{2}J_{2},\cdots ,r_{a}J_{a})$. Here by a linear form, we
mean a form of bi-degree (1,1). We have the following proposition.

\begin{prop}
\label{prop dim}
\begin{equation}
\dim \,N(m;r_{1}J_{1},r_{2}J_{2},\cdots ,r_{a}J_{a})\geq
(m_{1}+1)(m_{2}+1)-\sum_{j}\frac{r_{j}(r_{j}+1)}{2}  \label{g5}
\end{equation}%
with equality when $m_{i}\geq (\sum_{j}r_{j})-1,i=1,2$.
\end{prop}

\textit{Proof.} Without loss of generality, we consider a point $%
T=([0,1],[0,1])$. A form can be written as $F(z_{0},z_{1},y_{0},y_{1})=%
\sum_{i_{0}+i_{1}=m_{1},j_{0}+j_{1}=m_{2}}a_{I,J}z_{0}^{i_{0}}z_{1}^{i_{1}}y_{0}^{j_{0}}y_{1}^{j_{1}}
$ with $I=(i_{1},i_{2}),J=(j_{1},j_{2})$ are multi-indices. We call the
curve cut out by $F$ as $F$ as well. It has multiplicity at least $r$ at $T$
if and only if those coefficients with $i_{0}+j_{0}<r$ are zero. There are $%
1+2+\cdots +r=\frac{r(r+1)}{2}$ such coefficients, so such condition will
decrease the dimension at most $\frac{r(r+1)}{2}$. With equality if it is
the first condition to impose, and it can be less if several such conditions
are all assumed. Hence we have claimed inequality.

For the equality when $m_{i}$ is large, we put an induction on $%
t:=(\sum_{i}r_{i})-1$. If, say $m_{1}=1$, then this enforces $t=0$ or $1$.
If $t=0$, there would only be one $r_{i}$, and the dimension count above
will give the result since the redundancy in dimension decrease will not
arise. If $t=1$, for the same reason as before, we just need to consider $%
a=2,r_{1}=r_{2}=1$, so we are considering a curve passing through 2 distinct
points. This gives us 2 linearly independent linear equations on the
coefficients, which decreases the dimension of $N(m;r_{1}J_{1},r_{2}J_{2},%
\cdots,r_{a}J_{a})$ by 2.

Next we will consider cases where $m_{i}>1,t>1$. Suppose, firstly, each $%
r_{i}$ is 1. Every term $\frac{r_{1}(r_{1}+1)}{2}$ would be 1. By induction,
we only need to show that, every time we add a $J_{i}$, the dimension goes
down by 1. Set $N_{s}$ to be $N(m;J_{1},J_{2},\cdots ,J_{s})$. All we need
to do is to prove is that $N_{s}\neq N_{s-1}$. We choose linear forms $H_{i}$
passing through $J_{i}$ but not other $J_{j}$, another form $L_{0}$ not
passing all $J_{i}$ is used to complete the degree, we set $L_{0}$ to have
bi-degree $(m_{1}-s+1,m_{2}-s+1)$. Set $F=L_{1}L_{2}\cdots L_{s-1}L_{0}$,
and it will be in $N_{s-1}$ but not $N_{s}$, so $N_{s}\subsetneq N_{s-1}$.

Secondly, if $r=r_{1}>1$. Once again we set $T=J_{1}=([0,1],[0,1])$. In
order to decrease $t$, we consider:
\begin{equation}
N_{0}=N(m;(r-1)J_{1},r_{2}J_{2},\cdots ,r_{a}J_{a}).
\end{equation}%
We want to show that the dimension goes down at least $r$ when we come from $%
N(m;(r-1)J_{1},r_{2}J_{2},\cdots ,r_{a}J_{a})$ to $N(m;rJ_{1},r_{2}J_{2},%
\cdots ,r_{a}J_{a})$. This fact, the proved inequality together with the
induction hypothesis which gives the dimension of $%
N(m;(r-1)J_{1},r_{2}J_{2},\cdots ,r_{a}J_{a})$ will complete the induction.
To this end, we consider write $F\in N_{0}$ in the form $F=%
\sum_{i=0}^{r-1}a_{i}z_{0}^{i}y_{0}^{r-1-i}F_{i}+F^{\prime }$,~where $%
F^{\prime }$ contains those terms with greater sum of the degrees of $z_{0}$
and $y_{0}$. The multiplicity of $J_{1}$ is the smallest sum of the degrees
of $z_{0}$ and $y_{0}$. Define $N_{i},i\geq 1$ to be the subspace of $N_{0}$
satisfying $a_{0}=\cdots =a_{i-1}=0$. And notice that $N_{r}$ is the
subspace of $N_{0}$ that all $a_{i}$ mentioned vanishes, enforcing the
multiplicity of $J$ to be at least $r$. Hence we have a decreasing sequence
of spaces $N_{i}\supset N_{i+1},i=0,1,\cdots ,r-1$, and $%
N_{r}=N(m;rJ_{1},r_{2}J_{2},\cdots ,r_{a}J_{a})$. Consequently, similar to
the first case, all we need to show is that $N_{i}\neq N_{i+1},i=0,1,\cdots
,r-1$. And we will take this strengthened version of dimension count as a
part of the induction process.

We make an induction now. Define $W_{0}=N(m-(1,0);(r-2)J_{1},r_{2}J_{2},%
\cdots ,r_{a}J_{a})$. For $F\in W_{0}$, write $F=%
\sum_{i=0}^{r-1}b_{i}z_{0}^{i}y_{0}^{r-2-i}F_{i}+F^{\prime }$ and define $%
W_{k}$ to be the subspace of $W_{0}$ consisting of forms with $b_{j}=0,j<k$.
By induction, we have $W_{i}\supsetneq W_{i+1},i=0,1,\cdots ,r-2$, and $%
W_{r-1}=N(m-(1,0);(r-1)J_{1},r_{2}J_{2},\cdots ,r_{a}J_{a})$. So we can pick
$G_{i}$ such that $G_{i}\in W_{i},G_{i}\notin W_{i+1}$. Then $y_{0}G_{i}\in
N_{i}$ and $y_{0}G_{i}\notin N_{i+1}$. The case where the power of $y_{0}$
can not be increased can be treated by multiplying $z_{0}$: pick $G_{i}$ as
above, we know $z_{0}G_{r-2}\in N_{r-1},z_{0}G_{r-2}\notin N_{r}$. Hence $%
N_{i}\neq N_{i+1}$, and this completes the proof. {\hfill $\square $}\newline

With proposition \ref{prop dim} on dimension, we return to the proof of (\ref%
{eqn_g_04}), now the $N_{m}$ should be%
\begin{equation}
N_{m}=N(m;(r_{P}-1)P).
\end{equation}%
$P$ runs over those with $r_{P}\geq 2$. Using the exact sequence (\ref{g2})
we have%
\begin{equation}
l(E_{m})=(m_{1}+1)(m_{2}+1)-M-(m_{1}-d_{1}+1)(m_{2}-d_{2}+1).  \label{g7}
\end{equation}%
Then, using Riemann-Roch we hence have%
\begin{equation}
\begin{array}{lr}
g~=\deg \,E_{m}-l(E_{m})+1 &  \\
~~~=m_{1}d_{2}+m_{2}d_{1}-2M-((m_{1}+1)(m_{2}+1)-M-(m_{1}-d_{1}+1)(m_{2}-d_{2}+1))+1
&  \\
~~~=(d_{1}-1)(d_{2}-1)-M. &
\end{array}%
\end{equation}%
Hence we proved proposition \ref{prop_genus} in Section \ref%
{sec_genus_formula}.

\section{Topological Data of Subvarieties and Generalized Complete
Intersections}

\label{appendix_topological_data}

\renewcommand{\theequation}{B.\arabic{equation}} \setcounter{equation}{0} %
\renewcommand{\thethm}{B.\arabic{thm}} \setcounter{thm}{0} %
\renewcommand{\thelemm}{B.\arabic{lemm}} \setcounter{lemm}{0}

\vspace{1pt}

In this Appendix, we include supplementary details on determining the Hodge
numbers of the manifolds and using them as topological data to classify
those manifolds. We will explore the construction in \cite{Anderson:2015iia}%
. In this construction, $A$ is the ambient space, $M$ is the polynomial
intersection, and $X$ is the gCICY. In sum, we have $X\subset M\subset A$.
The main tools we will use are Hodge numbers and we compute them by
cohomology groups of sheaves. The methods in this Appendix are alternative
and complimentary to the approach in Section \ref%
{sec_spectral_sequence_approach} using spectral sequences of double
complexes.

\subsection{Adjunction Formula, Koszul Sequence and Euler Sequence}

To compute the cohomology groups of sheaves, we need the following short
exact sequences:

\begin{lemm}
(Adjunction Formula) For a divisor $D$ in the manifold $M$, we have
\begin{equation}
0\longrightarrow TD\longrightarrow TM|_{D}\longrightarrow
O_{M}(D)|_{D}\longrightarrow 0.
\end{equation}
\end{lemm}

\begin{lemm}
(Koszul sequence) Suppose $I_{D}$ is the ideal sheaf of the divisor $%
D\subset M$, we have
\begin{equation}
0\longrightarrow I_{D}\longrightarrow O_{M}\longrightarrow
O_{M}|_{D}\longrightarrow 0.
\end{equation}
\end{lemm}

Since the Euler sequence respects the direct sum, we have the following:

\begin{lemm}
(Euler sequence) For $A=A_{1}\times A_{2}$, $A_{1}=\mathbb{P}^{n_{1}}$, $%
A_{2}=\mathbb{P}^{n_{2}}$, we have
\begin{equation}
0\longrightarrow O_{A_{1}}\oplus O_{A_{2}}\longrightarrow O_{A}(1,0)^{\oplus
(n_{1}+1)}\oplus O_{A}(0,1)^{\oplus (n_{2}+1)}\longrightarrow
TA\longrightarrow 0.
\end{equation}
\end{lemm}

Tensoring the Koszul exact sequence with $TM$ and $O_{M}(D)$ respectively,
we obtain the following exact sequences:
\begin{equation}
0\longrightarrow O_{M}(-D)\otimes TM\longrightarrow TM\longrightarrow
TM|_{D}\longrightarrow 0.
\end{equation}%
\begin{equation}
0\longrightarrow O_{M}\longrightarrow O_{M}(D)\longrightarrow
O_{M}(D)|_{D}\longrightarrow 0.
\end{equation}

\subsection{Data of Subvariety $M$}

Now we choose the example $A=\mathbb{P}^{1}\times \mathbb{P}^{4}$,
\begin{equation}
M=%
\begin{matrix}
\mathbb{P}^{1} \\
\mathbb{P}^{4}%
\end{matrix}%
\begin{bmatrix}
3 \\
2%
\end{bmatrix}%
~~~\text{\textrm{and}}~~~X=%
\begin{matrix}
\mathbb{P}^{1} \\
\mathbb{P}^{4}%
\end{matrix}%
\begin{bmatrix}
3 &  & -1 \\
2 &  & 3{}%
\end{bmatrix}%
.  \label{gcicy1}
\end{equation}%
The reason we choose this example is that this is one of the simplest cases
with multiple columns and negative degrees, and it is a good demonstration
of computations of cohomology groups of sheaves and Hodge numbers. Also, it
is good for comparing with other computational methods for this example \cite%
{Anderson:2015iia,Garbagnati:2017rtb}. Since $X$ is a submanifold of $M$,
and $M$ is in turn a submanifold of $A$, we can use the adjunction formulas
in the short exact sequence form, iteratively. In order to derive the exact
sequence involving the data we need, we tensor appropriate sheaf with the
exact sequence we have. In particular, we can tensor it by tangent sheaves.

The Koszul sequence and adjunction formula for $M$ are:
\begin{eqnarray}
0 &\longrightarrow &O_{A}(-M)\longrightarrow O_{A}\longrightarrow
O_{A}|_{M}\longrightarrow 0.  \label{sequence_01} \\
0 &\longrightarrow &TM\longrightarrow TA|_{M}\longrightarrow
O_{A}(M)|_{M}\longrightarrow 0.  \label{sequence_02}
\end{eqnarray}%
Our first goal is to compute the Hodge numbers of $(M,TM)$. In this case, $%
I_{M}=O_{A}(-M)=O_{A}(-3,-2)$. To this end, we compute the cohomology groups
of $TA$ from the first sequence through the long exact sequence of
cohomology groups associated with this short exact sequence. Then, we tensor
the Koszul sequence with $TA$ to obtain the following exact sequence
\begin{equation}
0\longrightarrow TA\otimes O_{A}(-3,-2)\longrightarrow TA\longrightarrow
TA|_{M}\longrightarrow 0.  \label{7a}
\end{equation}%
Once we can compute the cohomology of $TA\otimes O_{A}(-3,-2)$, we would be
able to compute the cohomology of $TA|_{M}$ from this sequence. In order to
use the last sequence to compute the data of $TM$, we would need the data of
$O_{A}(3,2)|_{M}$. This can be obtained by tensoring Koszul sequence with $%
O_{A}(3,2)$:
\begin{equation}
0\longrightarrow O_{A}\longrightarrow O_{A}(3,2)\longrightarrow
O_{A}(3,2)|_{M}\longrightarrow 0.  \label{sequence_twisted_04}
\end{equation}%
The cohomology of $TA\otimes O_{A}(-3,-2)$ can be computed by K\"{u}nneth
formula. We consider the Bott vanishing theorem, as a special case of
Borel-Weil-Bott theorem.

\begin{thm}
(Bott Vanishing Theorem) \cite{Bott}
\begin{equation}
H^{q}(\mathbb{P}^{n},O_{\mathbb{P}^{n}}(k))=0,\mathrm{if}~\left\{ ~%
\begin{array}{l}
q\neq 0,0\leq k \\
\forall q,-n\leq k<0 \\
q\neq n,k\leq -(n+1)%
\end{array}%
\right. .
\end{equation}
\end{thm}

Another fact that we shall need is, when $k\geq 0$, $H^{0}(\mathbb{P}^{n},O_{%
\mathbb{P}^{n}}(k))$ is the space of degree-$k$ homogeneous polynomials, and
there is a perfect pairing:
\begin{equation}
H^{0}(\mathbb{P}^{n},O_{\mathbb{P}^{n}}(k))\times H^{n}(\mathbb{P}^{n},O_{%
\mathbb{P}^{n}}(-k-n-1))\longrightarrow \mathbb{C}.
\end{equation}%
So the dimensions of the two factors would be the same. So when $j\leq
-(n+1) $, we have a nonnegative $k$ such that $k=-j-(n+1)$. Using the above
duality, we have%
\begin{equation}
h^{0}(\mathbb{P}^{n},O_{\mathbb{P}^{n}}(k))=h^{n}(\mathbb{P}^{n},O_{\mathbb{P%
}^{n}}(j))=\binom{n+k}{n}=\binom{-j-1}{-j-n-1}.
\end{equation}

In sum, we have:

\begin{thm}
(Bott Formula) \cite{Okonek Schneider Spindler}
\begin{equation}
h^{q}(\mathbb{P}^{n},\Omega ^{p}(k))=\left\{
\begin{array}{lr}
\binom{k+n-p}{k}\binom{k-1}{p}\quad ~q=0,0\leq p\leq n,k>p &  \\
1\quad ~k=0,0\leq p=q\leq n &  \\
\binom{-k+p}{-k}\binom{-k-1}{n-p}~\quad q=n,0\leq p\leq n,k<p-n &  \\
0\quad ~otherwise &
\end{array}%
\right. .  \label{19}
\end{equation}%
Taking $p=0$, we will recover the formulas for $h^{q}(\mathbb{P}^{n},O_{%
\mathbb{P}^{n}}(k))$.
\end{thm}

To use this for $A$, which is the product of projective spaces, we need:

\begin{thm}
K\"unneth's Formula
\begin{eqnarray}
H^q(\mathbb{P}^{n_1}\times \mathbb{P}^{n_2}\times \cdots \times \mathbb{P}%
^{n_m},O(q_1,q_2,\cdots,q_m) )=\oplus_{k_1+k_2+\cdots+k_m=q} (H^{k_1}(%
\mathbb{P}^{n_1},O_{\mathbb{P}^{n_1}}(q_1)) \otimes  \notag \\
H^{k_2}(\mathbb{P}^{n_2},O_{\mathbb{P}^{n_2}}(q_2))\otimes \cdots\otimes
H^{k_m}(\mathbb{P}^{n_m},O_{\mathbb{P}^{n_m}}(q_m))).  \notag \\
\end{eqnarray}
\end{thm}

We also need the characterization of the tangent bundle using the hyperplane
bundle,%
\begin{equation}
T\mathbb{P}^{n}=\Omega ^{n-1}\otimes O(n+1).  \label{tangent_03}
\end{equation}

Now we proceed to compute the cohomology groups of $O_{A}(i,j)$. Notice that
those cohomology groups on the right hand side of K\"{u}nneth formula will
be nonzero only when $k_{i}=0,n_{i}$. Writing out the long exact sequence
associated to the Euler sequence and since the cohomology respects direct
sum, we have
\begin{equation}
\begin{array}{lr}
h^{k}(A,TA)=0,k\geq 1, &  \\
h^{0}(A,TA)=2\times 2+5\times 5-2=27, &  \\
h^{\ast }(A,TA)=(27,0,0,0,0,0). &
\end{array}%
\end{equation}

The second step shall be the calculation of $h^{k}(A,TA\otimes O_{A}(-3,-2))$%
. Denote the projection of $A$ onto the first and the second factor by $\pi
_{1},\pi _{2}$ respectively. We have:%
\begin{equation}
\begin{array}{l}
TA\otimes O_{A}(-3,-2) \\
=[\pi _{1}^{\ast }(T\mathbb{P}^{1}\otimes O_{\mathbb{P}^{1}}(-3))\otimes \pi
_{2}^{\ast }O_{\mathbb{P}^{4}}(-2)]\oplus \lbrack \pi _{1}^{\ast }O_{\mathbb{%
P}^{1}}(-3)\otimes \pi _{2}^{\ast }(T\mathbb{P}^{4}\otimes O_{\mathbb{P}%
^{4}}(-2))] \\
:=E_{1}\oplus E_{2}.%
\end{array}%
\end{equation}%
Now we substitute above formulas into the expression of $TA\otimes
O_{A}(-3,-2)$, to get $E_{1}=\pi _{1}^{\ast }O_{\mathbb{P}^{1}}(1)\otimes
\pi _{2}^{\ast }O_{\mathbb{P}^{4}}(-2).~$By K\"{u}nneth formula we know $%
h^{q}(\mathbb{P}^{4},O_{\mathbb{P}^{4}}(-2))$ will always be zero due to the
Bott formula. Hence $h^{q}(A,E_{1})=0.~$Then we turn to $E_{2}=\pi
_{1}^{\ast }O_{\mathbb{P}^{1}}(-3)\otimes \pi _{2}^{\ast }(\Omega
^{3}\otimes O_{\mathbb{P}^{4}}(3)).$ The $h^{q}(\mathbb{P}^{1},O(-3))$ is
nonzero only when $q=n=1$, being $3$. Then we consider $h^{q}(\mathbb{P}%
^{4},\Omega ^{3}\otimes O_{\mathbb{P}^{4}}(3))$. In this case we have $%
k=3,p=3,n=4$, they will never satisfy the conditions for nonzero ones in
Bott formula, hence those Hodge numbers will always vanish. So $%
h^{q}(A,E_{2})=0.~$ In conjunction with (\ref{7a}), the Hodge numbers of $%
TA|_{M}$ would be the same as that of $TA$, that is:
\begin{equation}
h^{\ast }(M,TA|_{M})=(27,0,0,0,0).
\end{equation}

To use the adjunction formula for $M$, we will need to calculate the Hodge
numbers of $O_{A}(3,2)|_{M}$. This can be obtained from the twisted Koszul
sequence (\ref{sequence_twisted_04}): The only nonzero Hodge number for the
cohomology groups of $O_{A}$ would locate at the 0-th cohomology group,
which is $1$. The only nonzero Hodge number of $O_{A}(3,2)$ would locate at
the 0-th cohomology group, which is $60$. So the only nonzero Hodge number
of $O_{A}(3,2)|_{M}$ would be at the 0-th cohomology group, being $60-1=59$.
Summing up, we have%
\begin{equation}
h^{\ast }(A,O_{A}(3,2))=(60,0,0,0,0,0),\ \ \ \ \ h^{\ast
}(A,O_{A}(3,2)|_{M})=(59,0,0,0,0,0).\
\end{equation}%
So by the long exact sequence associated to (\ref{sequence_02}), we see that
\begin{equation}
h^{\ast }(M,TM)=(0,32,0,0,0).  \label{17}
\end{equation}

\subsection{Data of Generalized Complete Intersection $X$}

Our essential goal would be the data of $X$. The adjunction formula of $X$
reads
\begin{equation}
0\longrightarrow TX\longrightarrow TM|_{X}\longrightarrow
O_{M}(-1,3)|_{X}\longrightarrow 0.  \label{8}
\end{equation}%
As long as the Hodge number of the latter two sheaves are both determined,
we would be able to determine the Hodge numbers of $(X,TX)$. First, for $%
O_{M}(-1,3)|_{X}$, we need the Koszul sequence for $X$:
\begin{equation}
0\longrightarrow O_{M}(1,-3)\longrightarrow O_{M}\longrightarrow
O_{M}|_{X}\longrightarrow 0.  \label{sequence_06}
\end{equation}%
Tensoring with $O_{M}(-1,3)$ we have:
\begin{equation}
0\longrightarrow O_{M}\longrightarrow O_{M}(-1,3)\longrightarrow
O_{M}(-1,3)|_{X}\longrightarrow 0.  \label{5}
\end{equation}%
Recall the information in (\ref{sequence_01}), we need the Hodge numbers of $%
O_{A}(-3,-2)$. All of these Hodge numbers will vanish due to K\"{u}nneth
formula and the fact that $h^{q}(\mathbb{P}^{4},O_{\mathbb{P}%
^{4}}(-2))=0,\forall q$. So the Hodge numbers of $O_{M}$ would be the same
as that of $O_{A}$:
\begin{equation}
h^{\ast }(M,O_{M})=(1,0,0,0,0).  \label{3}
\end{equation}%
For the Hodge numbers of $O_{M}(-1,3)$, we tensor (\ref{sequence_01}) with $%
O_{A}(-1,3)$ to obtain:
\begin{equation}
0\longrightarrow O_{A}(-4,1)\longrightarrow O_{A}(-1,3)\longrightarrow
O_{A}(-1,3)|_{M}\longrightarrow 0.  \label{6}
\end{equation}

The Hodge numbers of the first two sheaves can be computed through the Bott
formula. The cohomology groups of $O_{A}(-4,1)$ would be nonzero only for
the 1-st cohomology group for the first factor and the 0-th cohomology group
for the second factor. And its $h^{1}$ is $15$,
\begin{equation}
h^{\ast }(A,O_{A}(-4,1))=(0,15,0,0,0,0).
\end{equation}%
For the second sheaf, once again, we can observe that the first factor has
trivial cohomology groups, that is $h^{q}(\mathbb{P}^{1},O_{\mathbb{P}%
^{1}}(-1))=0,\forall q$, due to Bott formula, giving
\begin{equation}
h^{\ast }(A,O_{A}(-1,3))=(0,0,0,0,0,0).
\end{equation}%
So, combining the long exact sequence for (\ref{6}), we have
\begin{equation}
h^{\ast }(M,O_{M}(-1,3))=(15,0,0,0,0).
\end{equation}%
This result, (\ref{5}) and (\ref{3}) together, gives
\begin{equation}
h^{\ast }(X,O_{M}(-1,3)|_{X})=(14,0,0,0).
\end{equation}%
In spite of $h^{0}(A,O_{A}(-1,3))=0$, we see that $h^{0}(M,O_{M}(-1,3))\neq
0 $. This means that $X$ is an algebraic submanifold in $M$, although $X~$is
not a submanifold in $A$ defined by the global sections of line bundles on $%
A $.

So our focus now should be the cohomology groups of $TM|_{X}$. Tensoring the
sequence (\ref{sequence_06}) with $TM$, we have short exact sequence
involving $TM|_{X}$:
\begin{equation}
0\longrightarrow TM\otimes O_{M}(1,-3)\longrightarrow TM\longrightarrow
TM|_{X}\longrightarrow 0.  \label{18}
\end{equation}%
The Hodge numbers of $TM$ have been computed before, hence all we need to do
is to compute the Hodge numbers of $TM\otimes O_{M}(1,-3)$. The above
process used for $TM$ can be used here again after tensoring with $O(1,-3)$.
The Euler sequence will become:
\begin{equation}
0\longrightarrow O_{A}(1,-3)^{\oplus 2}\longrightarrow O_{A}(2,-3)^{\oplus
2}\oplus O_{A}(1,-2)^{\oplus 5}\longrightarrow TA\otimes
O_{A}(1,-3)\longrightarrow 0.  \label{11}
\end{equation}%
The anjunction formula will become:
\begin{equation}
0\longrightarrow TM\otimes O_{M}(1,-3)\longrightarrow (TA\otimes
O_{A}(1,-3))|_{M}\longrightarrow O_{M}(4,-1)\longrightarrow 0.  \label{13}
\end{equation}%
Firstly we compute the Hodge numbers of $O_{M}(4,-1)$. We tensor the Koszul
sequence (B.7) with $O_{A}(4,-1)$ to obtain:
\begin{equation}
0\longrightarrow O_{A}(1,-3)\longrightarrow O_{A}(4,-1)\longrightarrow
O_{M}(4,-1)\longrightarrow 0.
\end{equation}%
By the Bott formula applied to the first two sheaves, the second factor
which arise when we involve the K\"{u}nneth formula would always vanish,
hence they have identically trivial cohomology groups. Thus the long exact
sequence with respect to this short exact sequence enforces $O_{M}(4,-1)$ to
have trivial cohomology groups,
\begin{equation}
h^{\ast }(M,O_{M}(4,-1))=(0,0,0,0,0).
\end{equation}

What is left to be done is the cohomology of $TA\otimes O_{A}(1,-3)|_{M}$.
Tensor (\ref{sequence_01}) with $TA\otimes O_{A}(1,-3)$, we know the
following sequence is exact:%
\begin{equation}
0\longrightarrow TA\otimes O_{A}(-2,-5)\longrightarrow TA\otimes
O_{A}(1,-3)\longrightarrow (TA\otimes O_{A}(1,-3))|_{M}\longrightarrow 0.
\label{14}
\end{equation}%
The cohomology of the first two sheaves can be computed with the help of (%
\ref{tangent_03}). Similar to the computation on $TA\otimes O_{A}(-3,-2)$,
this sheaf will be:
\begin{equation}
\begin{array}{lr}
TA\otimes O_{A}(-2,-5) &  \\
=[\pi _{1}^{\ast }(T\mathbb{P}^{1}\otimes O_{\mathbb{P}^{1}}(-2))\otimes \pi
_{2}^{\ast }O_{\mathbb{P}^{4}}(-5)]\oplus \lbrack \pi _{1}^{\ast }O_{\mathbb{%
P}^{1}}(-2)\otimes \pi _{2}^{\ast }(T\mathbb{P}^{4}\otimes O_{\mathbb{P}%
^{4}}(-5))] &  \\
:=E_{3}\oplus E_{4}. &
\end{array}%
\end{equation}%
The first term is $E_{3}=\pi _{1}^{\ast }O_{\mathbb{P}^{1}}\otimes \pi
_{2}^{\ast }O_{\mathbb{P}^{4}}(-5)$. The Hodge numbers of the first factor
would be (1,0). That of the second sheaf would be obtained by Bott formula, $%
h^{4}(\mathbb{P}^{4},O_{\mathbb{P}^{4}}(-5))=1$. So the Hodge numbers of $%
E_{3}$ is $h^{\ast }(A,E_{3})=(0,0,0,0,1,0)$. The second part is $E_{4}=\pi
_{1}^{\ast }O_{\mathbb{P}^{1}}(-2)\otimes \pi _{2}^{\ast }(\Omega
^{3}\otimes O_{\mathbb{P}^{4}})$. The Hodge numbers of the first factor are $%
h^{\ast }(\mathbb{P}^{1},O_{\mathbb{P}^{1}}(-2))=(0,1)$. As for $\Omega
^{3}\otimes O_{\mathbb{P}^{4}}$, we have $k=0,p=3,n=4$, so the only
nontrivial Hodge number would be at $q=p=3$, being 1. Subsequently, $h^{\ast
}(\mathbb{P}^{4},\Omega ^{3}\otimes O_{\mathbb{P}^{4}})=(0,0,0,1,0)$. These
two sets of Hodge numbers give us$~h^{\ast }(A,E_{4})=(0,0,0,0,1,0)$. Since
the cohomology is compatible with direct sum, we see that
\begin{equation}
h^{\ast }(A,TA\otimes O_{A}(-2,-5))=(0,0,0,0,2,0).  \label{15}
\end{equation}

Likewise we consider%
\begin{equation}
\begin{array}{lr}
TA\otimes O_{A}(1,-3) &  \\
=[\pi _{1}^{\ast }(T\mathbb{P}^{1}\otimes O_{\mathbb{P}^{1}}(1))\otimes \pi
_{2}^{\ast }O_{\mathbb{P}^{4}}(-3)]\oplus \lbrack \pi _{1}^{\ast }O_{\mathbb{%
P}^{1}}(1)\otimes \pi _{2}^{\ast }(T\mathbb{P}^{4}\otimes O_{\mathbb{P}%
^{4}}(-3))] &  \\
:=E_{5}\oplus E_{6}. &
\end{array}%
\end{equation}%
We can simplify above expressions to be $E_{5}=\pi _{1}^{\ast }O_{\mathbb{P}%
^{1}}(3)\otimes \pi _{2}^{\ast }O_{\mathbb{P}^{4}}(-3)$ and $E_{6}=\pi
_{1}^{\ast }O_{\mathbb{P}^{1}}(1)\otimes \pi _{2}^{\ast }(\Omega ^{3}\otimes
O_{\mathbb{P}^{4}}(2))$. The second factor of $E_{5}$ has trivial cohomology
groups, hence $h^{\ast }(A,E_{5})=(0,0,0,0,0,0)$. For $E_{6}$, we know the
only nonzero Hodge number for $O_{\mathbb{P}^{1}}(1)$ would be $h^{0}(A,O_{%
\mathbb{P}^{1}}(1))=2$. For the second factor, we have $k=2,p=3,n=4$ in the
Bott formula, which gives trivial cohomology groups. Thus $h^{\ast
}(A,E_{6})=(0,0,0,0,0,0).~$Hence the second sheaf in sequence (\ref{14}) has
trivial cohomology groups:
\begin{equation}
h^{\ast }(A,TA\otimes O_{A}(1,-3))=(0,0,0,0,0,0).
\end{equation}%
By this and (\ref{15}), the long exact sequence associated with (\ref{14})
gives us:
\begin{equation}
h^{\ast }(M,(TA\otimes O_{A}(1,-3)|)_{M})=(0,0,0,2,0).  \label{16}
\end{equation}%
Together with (\ref{13}), it is now possible to determine the cohomology of $%
TM\otimes O_{M}(1,-3)$. Since the third sheaf has trivial cohomology groups,
the first two sheaves would have the same cohomology groups:
\begin{equation}
h^{\ast }(M,TM\otimes O_{M}(1,-3))=(0,0,0,2,0).
\end{equation}

In addition to this, with (\ref{17}), we are now ready to determine $h^{\ast
}(X,TM|_{X})$. We have a six-term long exact sequence for (\ref{18}),
\begin{equation}
\begin{array}{lr}
0\longrightarrow H^{1}(M,TM)_{32}\longrightarrow
H^{1}(X,TM|_{X})\longrightarrow 0 &  \\
\longrightarrow 0\longrightarrow H^{2}(X,TM|_{X})\longrightarrow
H^{3}(M,TM\otimes O_{M}(1,-3))_{2}\longrightarrow 0. &
\end{array}%
\end{equation}%
Here, the subscripts of the cohomology groups denote their dimensions. So we
have
\begin{equation}
h^{\ast }(X,TM|_{X})=(0,32,2,0).
\end{equation}%
Finally, we can compute the data of $TX$. A six-term long exact sequence for
(\ref{8}) will be
\begin{equation}
\begin{array}{lr}
0\longrightarrow H^{0}(X,O_{M}(-1,3)|_{X})_{14}\longrightarrow
H^{1}(X,TX)\longrightarrow H^{1}(X,TM|_{X})_{32} &  \\
\longrightarrow 0\longrightarrow H^{2}(X,TX)\longrightarrow
H^{2}(X,TM|_{X})_{2}\longrightarrow 0. &
\end{array}%
\end{equation}%
So we have%
\begin{equation}
h^{\ast }(X,TX)=(0,46,2,0).
\end{equation}%
Hence $h^{2,1}(X)=46,~h^{1,1}(X)=2,~h^{3,0}(X)=1~$and$\mathrm{~}h^{3}(X)=94$.%
$~$

In the above, we have used Serre duality and CY condition, from which we
know that
\begin{equation}
H^{1}(X,TX)=H^{2}(X,TX^{\ast })^{\ast }=H^{2,1}(X)^{\ast },
\end{equation}%
and
\begin{equation}
H^{2}(X,TX)=H^{1}(X,TX^{\ast })^{\ast }=H^{1,1}(X)^{\ast }.
\end{equation}



\end{document}